\documentclass{article}
\usepackage{amsmath, amsthm, amsfonts}
\usepackage[margin=1.25in]{geometry}
\usepackage{amsmath}
\usepackage{amssymb}
\usepackage{IEEEtrantools}
\usepackage{framed}
\usepackage[small]{titlesec}
\usepackage{fancyhdr}
\usepackage{cases}
\usepackage{xcolor}
\usepackage[all]{xy}

\usepackage{listings}
\usepackage{color} 
\definecolor{mygreen}{RGB}{28,172,0} 
\definecolor{mylilas}{RGB}{170,55,241}

 \usepackage{verbatim}

\usepackage{graphicx}
\usepackage{caption}
\usepackage{subcaption}
\usepackage{dblfloatfix}
\usepackage{float}

\usepackage[numbers,sort&compress]{natbib}
\usepackage{color}   
\usepackage{hyperref}
\hypersetup{
    colorlinks=true, 
    linktoc=all,     
    linkcolor=blue,  
}

\renewcommand\thesection{\Roman{section}}
\renewcommand\thesubsection{\Alph{subsection}}

\titleformat{\section}	
{\normalfont\bfseries\fillast}
{\bfseries \small {\thesection}.}
{1ex minus -10ex}
{\small}
\titlespacing{\section}
{3pc}{*3}{*2}[3pc]

\titleformat{\subsection}
{\normalfont\bfseries\fillast}
{\bfseries \small {\thesubsection}.}
{1ex minus -10ex}
{\small}
\titlespacing{\section}
{3pc}{*3}{*2}[3pc]

\titleformat{\subsubsection}
{\normalfont\bfseries\fillast}
{\bfseries \small {\thesubsubsection}}
{1ex minus -10ex}
{\small}
\titlespacing{\section}
{3pc}{*3}{*2}[3pc]

\theoremstyle{definition}

\theoremstyle{remark}

\title{Results}
\date{\vspace{-5ex}}

\begin{document}

\pagestyle{myheadings}

\title{A Toolkit For Steady States of Nonlinear Wave Equations: \\
Continuous Time Nesterov and Exponential Time Differencing Schemes}
\author{C.B. Ward$^1$, N. Whitaker$^1$, I.G. Kevrekidis$^2$ and
  P.G. Kevrekidis$^1$\\ 
$^1$ Department of Mathematics and Statistics, University of Massachusetts\\
Amherst, MA 01003, USA\\
$^2$  Department of Chemical and Biological Engineering, Princeton University \\
  Princeton, NJ 08544
}

\maketitle


\begin{abstract}
  Several methods exist for finding ground (as well as excited)
  states of nonlinear waves equations. In this paper we first
  introduce two modifications of the so-called accelerated imaginary-time
  evolution method (AITEM). In our first modification, 
  time integration of the underlying  gradient flow is done using exponential
  time differencing instead of using more standard methods. 
  In the second modification, we present a generalization of the gradient flow model, motivated by the work of Nesterov, as well as that of Candes and
  collaborators. Additionally, we consider combinations of these
  methods with the so-called spectral renormalization scheme.
  Finally, we apply these techniques to the
  so-called Squared Operator Method, enabling 
  convergence to excited states. 
  Various examples are shown to illustrate the effectiveness of these
  new schemes,
  comparing them to standard ones established in the literature. In most cases,
  we find significant reductions in the number of iterations needed to
  reach convergence.
\end{abstract}

\section{Introduction}

In models stemming from nonlinear optics and atomic physics, it is
customary to seek a Hamiltonian description of the dynamics, 
e.g., for the envelope of optical pulses or for the wavefunction of 
quantum systems that follows a Schr{\"o}dinger type partial
differential equation. If, in addition, nonlinear effects
are contributing, e.g., either because of the so-called Kerr effect
in optics~\cite{hasegawa,kivshar} or because of the mean-field
interaction of bosonic 
atoms~\cite{GP,becbook1,becbook2,rcg:BEC_BOOK,rcg:BEC_book2}, then
the prototypical model becomes the nonlinear Schr{\"o}dinger
(NLS) equation.
The NLS
\cite{ablowitz,abl2,ablowitz1,sulem,chap01:bourgain}
is a dispersive nonlinear partial differential equation (PDE)
that has been essential in understanding
some of the most groundbreaking results in the physics of such systems.
Additional areas of application include,
but are not limited to  Langmuir waves in plasmas \cite{zakh1,zakh2},  deep water and freak/rogue waves, 
\cite{benjamin,onofrio},
as well as more broadly in fluid mechanics~\cite{infeld}.

In its canonical form, the equation reads:
\begin{equation}
\begin{array}{rcl}
i \partial_t u=-\frac{1}{2}\nabla^2 u + g |u|^2 u,
\label{nls}
\end{array}
\end{equation}
where $u$ is the complex field and $g$ is a constant.
Physically, $u$ may represent the
envelope of the electric field in optics, the amplitude of
water waves or the wavefunction of a Bose-Einstein condensate (BEC) and
is the main object
whose spatio-temporal evolution we are
interested in probing.
Very commonly in the above areas, we are interested in
identifying standing wave solutions of Eq.~(\ref{nls})
in the form: $u(x,t)=\psi(x) e^{-i \mu t}$ which, in turn,
leads to the time-independent form of the equation:
\begin{IEEEeqnarray}{rcl}
  \nabla^2 \psi -V(x) \psi + \sigma |\psi|^2 \psi + \mu \psi  &\;=\;& 0.
  \label{eqn1}
\end{IEEEeqnarray}
The parameter $\mu$ is associated to the frequency of the
solution and is referred to as the propagation constant
in optics or the chemical potential in atomic BECs.
This steady state problem constitutes a subject of wide exploration, to which a
broad and diverse number of studies has been devoted.
Both the ground and the
excited states in this elliptic, nonlinear PDE problem are of interest.
It should
be noted that given the importance of the subject entire books
have been dedicated to the analysis of associated numerical
methods~\cite{yangbook}.

Our aim in the present work is to add some useful twists
to this extensive literature, based on recent computational
developments in other areas (including the time stepping of
ordinary and partial differential equations, and the development
of schemes relevant for the iterative convergence of functional
extremization). Our main contribution is to propose iterative schemes, based on 
the continuous time variant of Nesterov's method \cite{5,4,MJ}, for finding stationary states of  Eq.~(\ref{eqn1}).
The structure of our presentation is as follows. Given the extensive
literature on the subject related to the NLS model, we start by
presenting in section II
some of the most popular methods that do {\it not} resort
to the use of the Jacobian (i.e., Newton-type methods); the latter,
and accelerated variants thereof,
merit their own independent examination that is deferred to a future
stage. Then, we present in section III
our proposed ``twists'' based on the above
recent computational developments and their implementation
in Eq.~(\ref{eqn1}). In section IV we compare the results of
the newly proposed variants with the more standardly used
methods. Finally, in section V, we summarize our findings and
present some challenges for the future.

\section{Earlier Methods for Calculating Ground States}
In this section we discuss two among the most widely used,
previously developed methods, AITEM~\citep{2} and the
Spectral Renormalization method~\citep{3}, for identifying ground states of
the steady state problem within the NLS equation. 
\subsection{AITEM}
\noindent

Eq.~(\ref{eqn1}) can be recast in the variational form

\begin{IEEEeqnarray}{rcl}\label{VP}
\min_{\psi } \int | \nabla \psi |^2 +V(x)| \psi |^2 - \frac{\sigma }{2} | \psi |^4 \; dx & \qquad \text{subject to} \qquad & \int | \psi |^2  \; dx =P,
\end{IEEEeqnarray}
where the first integral is the field-theoretic energy $E(\psi)$
of the system
and the second integral fixes the number of particles  (in
the atomic case) or the power --hence the symbol-- in the optical
case to be $P$~\footnote{It is worth noting that while in the analysis below 
  we explore the cubic nonlinearity for concreteness,
  our considerations are, in principle, expected to apply equally well to
  more general nonlinearities.}.

Using a Lagrange multiplier, we can directly incorporate the
relevant constraint. The resulting gradient flow is then given by 
\begin{IEEEeqnarray}{rcl}\label{AITEM}
\dot{\psi} &\;=\;&\nabla^2 \psi -V(x) \psi + \sigma |\psi|^2 \psi + \mu \psi.
\end{IEEEeqnarray}
In tradional variational problems, $P$ is typically known and $\mu$ unknown i.e. the constraint is a given but the Lagrange multiplier must be identified.
In this case, we can let $\mu=\mu(t)$ be a function of time such that
$\mu(t)$ converges to the true value of the Lagrange multiplier as
$t \to \infty$. One such choice of $\mu(t)$ was given by
Yang and Lakoba~\citep{2} as
\begin{equation}
\mu(t) = - \frac{ \langle L \psi , \psi \rangle}{\langle \psi , \psi \rangle}
\end{equation}
where the inner products represent the standard $L^2$ inner product and $L=\nabla^2 \psi -V(x) \psi + \sigma |\psi|^2 \psi$; if one thinks of $\mu$ as an eigenvalue then this is the standard Rayleigh quotient. Because $\psi =0$ is always a solution of the NLS, one must still include the constraint $\int |\psi|^2 \; dx =P$ to ensure the evolution does not go to the trivial solution. If one
applies, say, the standard Euler method to \eqref{AITEM} and also adds a preconditioner $M$, then one gets the AITEM scheme:
\begin{IEEEeqnarray}{rcl}
M&\; = \;& c-\nabla^2 \nonumber \\[.2cm]
\mu_n & \; = \; & \frac{\langle M^{-1}L \psi_n, \psi_n \rangle}{\langle M^{-1}\psi_n, \psi_n\rangle} \nonumber \\[.2cm]
\tilde{\psi}_{n+1} & \; = \; & \psi_n - M^{-1}(L \psi_n + \mu_n \psi_n) \Delta t\\[.2cm]
\psi_{n+1} &\;=\;& \tilde{\psi}_{n+1} \sqrt{\frac{P}{\langle \tilde{\psi}_{n+1},\tilde{\psi}_{n+1} \rangle}}. \nonumber
\end{IEEEeqnarray}
The fourth equation ensures that the number of particles (the constraint $\int |\psi|^2 \; dx =P$) is satisfied after each iteration. We remark that the parameter $c$ is a positive number which must be chosen a priori.

\subsection{Spectral Renormalization}
An alternative method applicable to the NLS for
general nonlinearity $N$ is the so-called spectral renormalization method,
addressing problems of the form:
\begin{IEEEeqnarray*}{rcl}
  \nabla^2 \psi -V(x) \psi + N(|\psi |^2)\psi + \mu \psi  &\;=\;& 0
  \label{eqn2}
\end{IEEEeqnarray*}
Unlike before, here we think of $\mu$ as a fixed constant. If we take the Fourier transform (denoted by ${\cal F}$) of this equation we obtain
\[-|\textbf{k}|^2 \hat{\psi} + \mathcal{F} [- V(x) \psi + N(|\psi |^2)\psi] + \mu \hat{\psi} = 0\]
and solving for $\hat{\psi}$ yields
\[\hat{\psi} = \frac{\mathcal{F} [- V(x) \psi + N(|\psi |^2)\psi]}{|\textbf{k}|^2 - \mu}\]
Thinking of this as a fixed point iteration method
\[\hat{\psi}_{n+1} = \frac{\mathcal{F} [- V(x) \psi_n + N(|\psi_n |^2)\psi_n]}{|\textbf{k}|^2 - \mu}\]
we might expect this to converge to a ground state. However numerical experiments have shown that it tends to converge to zero or diverge without bound.

To get around this problem, Ablowitz and Musslimani \citep{3} suggested that one should include a renormalization factor $\lambda$, which is determined by the iteration procedure itself. Letting $\psi = \lambda \phi$, $\hat{\psi} = \lambda \hat{\phi}$, plugging these into the NLS equation, and repeating gives
\[\hat{\phi} = \frac{\mathcal{F} [- V(x) \phi + N(|\lambda \phi |^2)\phi]}{|\textbf{k}|^2 - \mu}\]
If we now multiply the previous equation by $\hat{\phi}$ and integrate we get an algebraic condition on $\lambda$:
\[\langle \hat{\phi} , \hat{\phi} \rangle - \langle \hat{\phi} , \frac{\mathcal{F} [- V(x) \phi + N(|\lambda \phi |^2)\phi]}{|\textbf{k}|^2 - \mu} \rangle = 0\]
Since $\lambda$ is just a scalar, we see that it is determined by the above equation. We then have the scheme:
\begin{IEEEeqnarray*}{rcl}
0 &\;=\;& \langle \hat{\phi}_n , \hat{\phi}_n \rangle - \langle \hat{\phi}_n , \frac{\mathcal{F} [-V(x) \phi_n + N(|\lambda_n \phi_n|^2)\phi_n]}{|\textbf{k}|^2 - \mu} \rangle  \nonumber \\[.2cm]
\hat{\phi}_{n+1}&\;=\;& \frac{\mathcal{F} [-V(x) \phi_n + N(|\lambda_n \phi_n|^2)\phi_n]}{|\textbf{k}|^2 - \mu}
\end{IEEEeqnarray*}

One drawback of the scheme as written is that if $\mu$ is not negative then the iteration leads to division by zero. 
In~\cite{3}, it was thus suggested that the term $r \psi$ be added and subtracted to the NLS equation; if one then repeats the argument, 
a scheme where division by zero does not occur can be devised. This scheme, the Spectral Renormalization method, is given by 
\begin{IEEEeqnarray}{rcl}
0 &\;=\;& \langle \hat{\phi}_n , \hat{\phi}_n \rangle - \langle \hat{\phi}_n , \frac{(r + \mu)\hat{\phi}_n}{r+|\textbf{k}|^2} + \frac{\mathcal{F} [-V(x) \phi_n + N(|\lambda_n \phi_n|^2)\phi_n]}{r+|\textbf{k}|^2} \rangle  \nonumber \\[.2cm]
\hat{\phi}_{n+1}&\;=\;&  \frac{(r + \mu)\hat{\phi}_n}{r+|\textbf{k}|^2} + \frac{\mathcal{F} [-V(x) \phi_n + N(|\lambda_n \phi_n|^2)\phi_n]}{r+|\textbf{k}|^2}.
\end{IEEEeqnarray}
where $r$ is some positive parameter which must be chosen before the iteration begins.

\section{Proposed Twists}
In this section we propose a number of
modifications and extensions of AITEM and Spectral Renormalization.

\subsection{Exponential Time Differencing}
The first of these new methods is simply a different way of
time-stepping the gradient flow equation. Namely, using the first-order exponential time differencing scheme \citep{7,8} instead of Euler's method. 

More specifically, consider Eq.~\eqref{AITEM} again. By taking the Fourier transform of both sides we arrive at
\[\hat{\psi}_t=-|\textbf{k}|^2 \hat{\psi} + \mathcal{F} [- V(x) \psi + \sigma |\psi |^2\psi] + \mu \hat{\psi} \]
Applying the first order exponential time differencing scheme to this equation we get
\[\hat{\psi}_{n+1} = e^{-|\textbf{k}|^2h}\hat{\psi}_n + \frac{e^{-|\textbf{k}|^2h} - 1}{-|\textbf{k}|^2}[\mathcal{F} [- V(x) \psi _n+ \sigma |\psi_n |^2\psi_n] + \mu \hat{\psi}_n]\]
However, some care is needed in dealing with the term $\frac{e^{-|\textbf{k}|^2h} - 1}{-|\textbf{k}|^2}$ so that division by zero and catastrophic cancellation do not occur. We refer the reader to the  insightful work of Kassam and Treffethen \cite{8} in which they propose to use the Cauchy integral formula to calculate this expression and include a Matlab code for implementing this at the end.

Now, we have to impose the constraint $\int |\psi|^2 dx = P$.
We proceed in a similar fashion as AITEM:
\begin{IEEEeqnarray}{rcl}
\mu_n & \; = \; & \frac{\langle L \hat{\psi}_n, \hat{\psi}_n \rangle}{\langle \hat{\psi}_n, \hat{\psi}_n\rangle} \nonumber \\[.2cm]
\tilde{\hat{\psi}}_{n+1} & \; = \; & e^{-|\textbf{k}|^2h}\hat{\psi}_n + \frac{e^{-|\textbf{k}|^2h} - 1}{-|\textbf{k}|^2}[\mathcal{F} [- V(x) \psi _n+ \sigma |\psi_n |^2\psi_n] + \mu_n \hat{\psi}_n] \\[.2cm]
\hat{\psi}_{n+1} &\;=\;& \tilde{\hat{\psi}}_{n+1} \sqrt{\frac{P}{\langle \tilde{\hat{\psi}}_{n+1},\tilde{\hat{\psi}}_{n+1} \rangle}} \nonumber
\end{IEEEeqnarray}
where $L\hat{\psi}= -|\textbf{k}|^2 \hat{\psi} + \mathcal{F} [- V(x) \psi+ \sigma |\psi|^2\psi]$ and $h=\Delta t$, the proposed effective time step. We will refer to this scheme as ETD for the remainder of the paper. Our main motivation for proposing this scheme is that it does not need a preconditioner like that in AITEM; in some sense, Duhamel's formula itself --incorporating
the integration of the Laplacian term-- is a preconditioner. We also expect that if the potential stiffness is due to the Laplacian term,
then this method should perform quite well.

If the stiffness is instead concentrated in the term $V(x) \psi$ then we expect ETD and AITEM to do far more poorly. In such a case, we propose that $V(x) \psi$ 
should be considered the linear part and not $\nabla^2 \psi$. Before proceeding, we remark that if one does exponential time differencing 
in physical space then it is difficult to compute the operator
$e^{\nabla^2 h}$. Moreover, 
in Fourier space it is difficult to separate $\hat{\psi}$ in $\mathcal{F}[V(x) \psi]$ from the potential; what this implies computationally is that one must 
choose between letting the Laplacian {\it or} the potential to be included
in the linear part. 

Now, staying in physical space and performing exponential time differencing
based on the potential gives
\begin{IEEEeqnarray}{rcl}
\mu_n & \; = \; & \frac{\langle L \psi_n, \psi_n \rangle}{\langle \psi_n, \psi_n\rangle} \nonumber \\[.2cm]
\tilde{\psi}_{n+1} & \; = \; & e^{-V(x) h} \psi_n + \frac{e^{-V(x) h} - 1}{-V(x)}[\nabla^2 \psi _n+ \sigma |\psi_n |^2\psi_n + \mu_n \psi_n]\\[.2cm]
\psi_{n+1} &\;=\;& \tilde{\psi}_{n+1} \sqrt{\frac{P}{\langle \tilde{\psi}_{n+1},\tilde{\psi}_{n+1} \rangle}} \nonumber
\end{IEEEeqnarray}
where, again, the term $\frac{e^{-V(x) h} - 1}{-V(x)}$ must be interpreted appropriately. We will refer to this scheme as ETDV.

\subsection{Continuous Time Nesterov}
Consider the variational problem of minimizing the function $F(x)$; here we are considering $F$ to be a function and not a functional. To solve this problem, one method is of course to use gradient descent. However, if $F$ is sufficiently
``ill-behaved'' we do not expect that gradient descent will converge easily.
As an alternative, Su, Boyd, and Candes \citep{4} were able to formulate a second order ODE which in some sense generalizes gradient descent:
\[\ddot{x} + \frac{3}{t}\dot{x} + \nabla F(x) =0\]
As discussed in their paper, this ODE is actually a continuous version of Nesterov's (discrete) mirror descent \citep{5}. Henceforth, we will refer to this
scheme as continuous time Nesterov (CTN).

Two major differences occur between CTN and gradient descent. The first, and crucial one (since it will also enable the second as we will see), is that CTN is a second order ODE. Roughly speaking, this means that the acceleration vector, and NOT the velocity vector, points in the direction that the field is decreasing fastest (at least for large $t$). This is similar to a particle moving in
the force field of a 
potential i.e. a related way of envisioning this ODE is to say that the particle has been given mass and has a time-dependent dissipation
on which we now comment. The second major difference is the dissipation term $\frac{3}{t}\dot{x}$; thinking of a particle in a  potential, we see that this term has the effect of damping the energy/momentum.
However, this damping is tuned to be large at the initial time, when
presumably the particle is far from the equilibrium while it
decreases the closer that one (hopefully) gets to the relevant fixed point.
This term is, thus, responsible for the actual convergence of the method to minima of $F$. With too little damping the method will only oscillate around the minima but with too much damping the method could be terribly inefficient. 

In the work of \cite{4}, the authors suggest using a second-order center difference scheme for approximating the second derivative and a first order backward difference scheme for approximating the first derivative. Doing this and rearranging the dynamical evolution equation gives the scheme 
\[x_{n+1} = (2- \frac{3}{n})x_n - (\Delta t)^2 \nabla F(x_n) - (1-\frac{3}{n})x_{n-1}\]
where we have let $t=n\Delta t $. 

We remark in passing that, as was proven in \citep{5}, this scheme enjoys linear convergence, provided $F$ is strongly convex. 

\subsection{Accelerated Continuous Time Nesterov}

A principal contribution of the present work
is to propose and illustrate the relevance of applying
CTN not just to functions but to (field-theoretic) functionals; as far as we
know, this application of CTN as a means of finding steady state solutions of
a PDE has not been previously considered. 

Returning to the variational problem \eqref{VP}, we see that CTN takes the form
\[\ddot{\psi} + \frac{3}{t}\dot{\psi} -( \nabla^2 \psi -V(x) \psi + \sigma |\psi|^2 \psi + \mu \psi) =0\]
where we have  included the Lagrange multiplier $\mu$, and 
abused the overdot notation in this field-theoretic context to signify
partial derivative with respect to $t$. Discretizing this as before, we arrive at
\[\psi_{n+1} = (2- \frac{3}{n})\psi_n + (\Delta t)^2 (\nabla^2 \psi -V(x) \psi + \sigma |\psi|^2 \psi + \mu \psi) - (1-\frac{3}{n})\psi_{n-1}.\]

Since the dissipation term controls the convergence properties to a high degree, both in the work of~\cite{4} and in that of~\citep{6}, much effort has
been invested in trying to optimize it. In particular,
it is proposed to reset time $t$ at appropriate points in the evolution so that CTN is always sufficiently damped; again, when $t$ is small there is a large amount of damping. Such a variant is the gradient restarting scheme, whereby time is reset to one
when the angle between $-\nabla F(x)$ and $\dot{x}$ is greater than $90$ degrees AND a prespecified amount of time $t_{res}$ has elapsed:
\begin{IEEEeqnarray*}{rcl}
\langle \nabla F(x) , \dot{x} \rangle &>& 0\\
t&\geq &t_{res}.
\end{IEEEeqnarray*}
If we include gradient restarting into the above descritization we get
\begin{equation*}
\psi_{n+1} = (2- \frac{3}{\tilde{n}})\psi_n + (\Delta t)^2 (\nabla^2 \psi_n -V(x) \psi_n + \sigma |\psi_n|^2 \psi_n + \mu \psi_n) - (1-\frac{3}{\tilde{n}})\psi_{n-1},
\end{equation*}
where $\tilde{n}$ starts at one and increases by one after each iteration; once the restart condition
\begin{IEEEeqnarray}{rcl}
\langle \nabla^2 \psi_n -V(x) \psi_n + \sigma |\psi_n|^2 \psi_n + \mu \psi_n \; \; , \; \; \psi_{n+1} - \psi_n \rangle &>&0 \nonumber \\
n &\geq& n_{res}
\end{IEEEeqnarray}\label{eq:RestartReal}
is met, $\tilde{n}$ is reset to one and the process repeats.

If we include a preconditioner $M$ and recall that we must normalize after each iteration, then the full method can be written as
\begin{IEEEeqnarray}{rcl}\label{ACTN}
M &=& c - \nabla ^2 \nonumber \\[.2cm]
\mu_n & \; = \; & \frac{\langle L \psi_n, \psi_n \rangle}{\langle \psi_n, \psi_n\rangle} \nonumber \\[.2cm]
\tilde{\psi}_{n+1} &=& (2- \frac{3}{\tilde{n}})\psi_n + (\Delta t)^2 M^{-1} (\nabla^2 \psi_n -V(x) \psi_n + \sigma |\psi_n|^2 \psi_n + \mu_n \psi_n) - (1-\frac{3}{\tilde{n}})\psi_{n-1}\\[.2cm]
\psi_{n+1} &\;=\;& \tilde{\psi}_{n+1} \sqrt{\frac{P}{\langle \tilde{\psi}_{n+1},\tilde{\psi}_{n+1} \rangle}} \nonumber
\end{IEEEeqnarray}
where, again, $\tilde{n}$ is chosen via gradient restarting. We shall refer to this scheme as Accelerated Continuous Time Nesterov (ACTN), in the fashion of AITEM.

We remark that the convergence rate of this method is unknown to us. While CTN was proven to converge linearly under strong convexity, no convergence proof is known to us of CTN with gradient restarting (though Su, Boyd, and Candes prove something similar). With the inclusion of the preconditioner and particle number normalization, it is not clear what convergence 
speed should be expected. To that end, the numerical experiments below suggest the ACTN will, generically, converge linearly.

Lastly, we note that if we fix $\mu$ in the NLS equation then AITEM, ACTN, and ETD can all be renormalized via a straightforward procedure that
we present in the Appendix. We denote these as Renormalized AITEM (AITEMRe), Renormalized ACTN (ACTNRe), and Renormalized ETD (ETDRe). We also mention that, in principle, this procedure can be done for more general constraints.

\section{Computational Results}
We now present the results of the realization of the proposed
methods for fundamental as well as excited steady states of the
one- and two-dimensional NLS equation with different types of trapping
potentials. Each example has a comparison with AITEM and Spectral Renormalization to give a reference point.

\subsection{Ground States in 1D}
Unless otherwise mentioned, we take the spatial domain to be $[-12,12]$. For all methods except ETDV, spatial descritization is done in Fourier space via the Discrete Fourier Transform (DFT) with 128 points. ETDV is discretized in physical space using finite differences with 128 points.
The initial condition used for all examples is $\psi_0 = A e^{-x^2}$, where $A$ was chosen so that the power is five.

We also want to emphasize that we are computing AITEM, AITEMRe, ACTN, and ACTNRE in Fourier space. To be precise, we first take the DFT of the given equation (gradient descent or CTN) and then we apply the given iteration procedure to this equation. Doing it this way, the cost of one iteration of each of AITEM and ACTN involves only one FFT and one IFFT; it also makes the computation of $M^{-1}$ very cheap. The renormalized methods will cost slightly more depending on the equation. For example, the scheme \eqref{AITEMRE}, see the appendix, will cost two FFT's and one IFFT per iteration.

\begin{figure}[]
\centering
    \begin{subfigure}[c]{0.45\textwidth}
        \includegraphics[width=\textwidth, keepaspectratio=true]{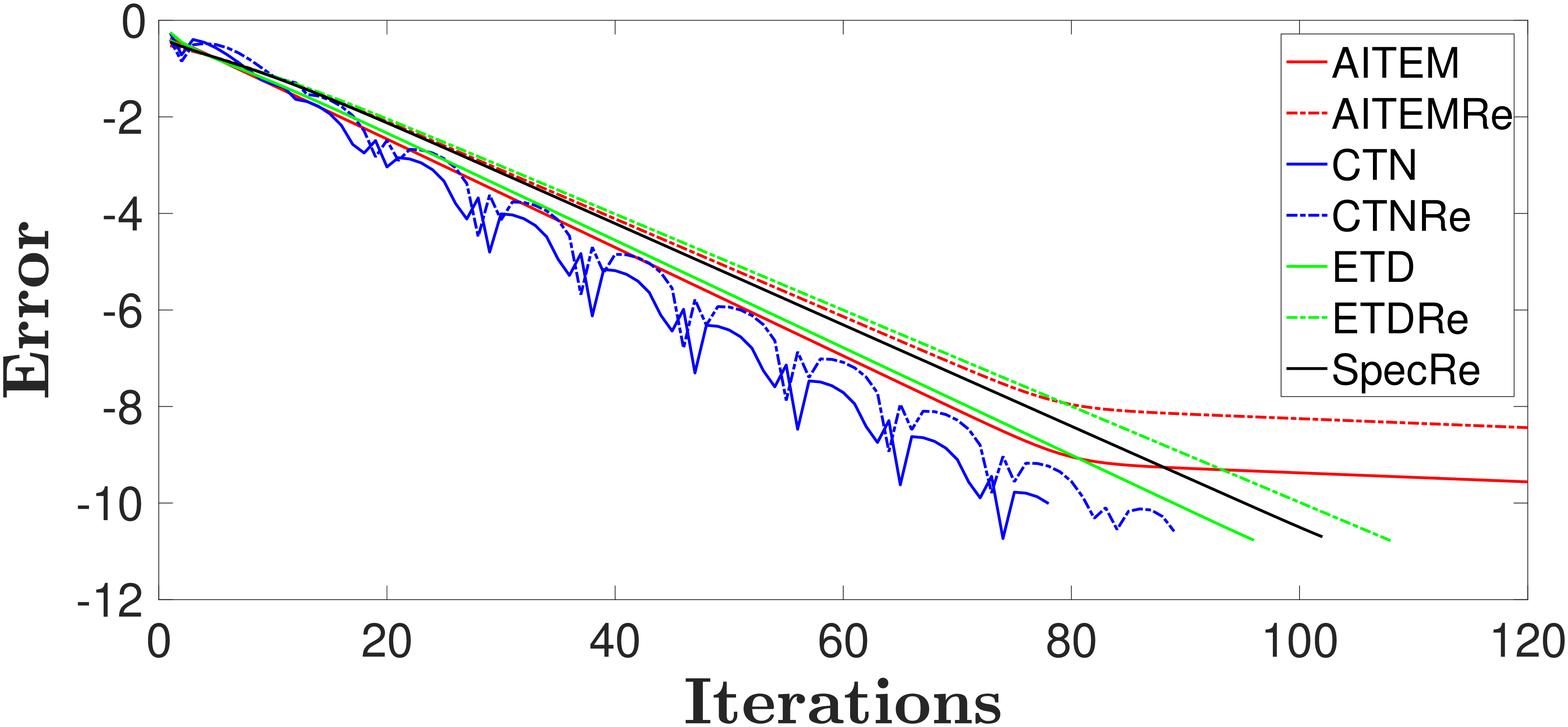}
        \caption*{(a) $V(x) =0.1x^2$ , $\sigma=-1$}
    \end{subfigure}
    \quad
        \begin{subfigure}[c]{0.42\textwidth}
        {\footnotesize
        \begin{tabular}{l*{5}{c}}
Scheme              & $\Delta t$ & $c$ & $\tilde{n}$ & $r$ & Iterations\\
\hline \\[.05cm]
AITEM 				& .55 & 3 & - & - &   251\\[.05cm]
AITEMRe          & .55 & 3 & - & -  & 370\\[.05cm]
ACTN          		& .85 & 3 & 9 & -  & 78\\[.05cm]
ACTNRe    		& .85 & 3 & 9 & - & 89\\[.05cm]
ETD   				& .16 & - & - & - & 96\\[.05cm]
ETDRe   			& .16 & - & - & - & 108\\[.05cm]
SpecRe   			& - & - & - & 5.7 & 102\\[.05cm]
ETDV   				& .017 & - & - & - & 960\\[.05cm]
\end{tabular}
}
        \caption*{(b) $\mu = 1.1848$}
            \end{subfigure}
     \qquad
    \begin{subfigure}[c]{0.45\textwidth}
        \includegraphics[width=\textwidth, keepaspectratio=true]{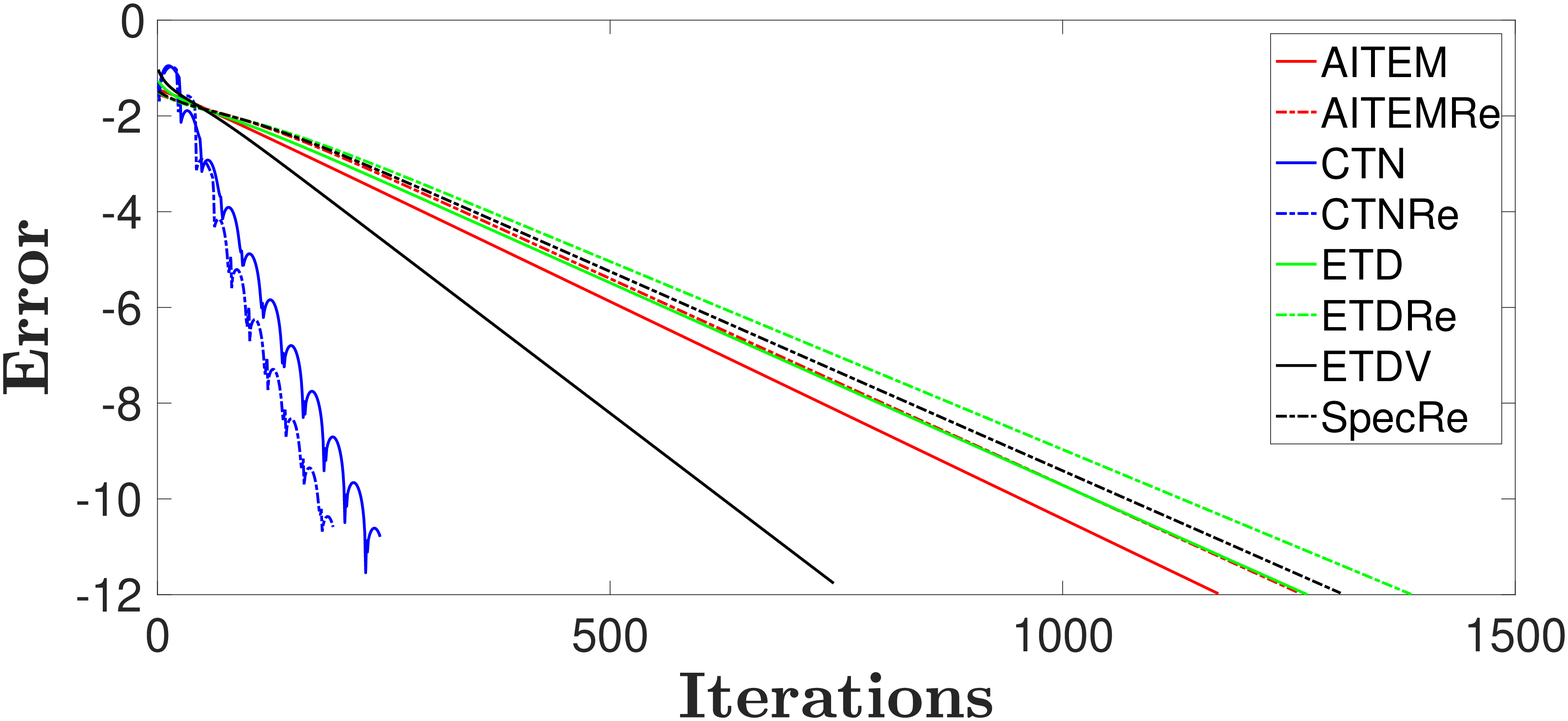}
        \caption*{(c) $V(x) =0.01x^4 + 0.02x^2$ , $\sigma=-1$}
    \end{subfigure}
        \quad
        \begin{subfigure}[c]{0.42\textwidth}
        {\footnotesize
        \begin{tabular}{l*{5}{c}}
Scheme              & $\Delta t$ & $c$ & $\tilde{n}$ & $r$ & Iterations\\
\hline \\[.05cm]
AITEM 				& .052 & 4 & - & - &  1172 \\[.05cm]
AITEMRe          & .14 & 12 & - & -  & 1260 \\[.05cm]
ACTN          		& .3 & 5 & 23 & -  & 246\\[.05cm]
ACTNRe    		& .35 & 6 & 20 & - & 194\\[.05cm]
ETD   				& .01 & - & - & - & 1270\\[.05cm]
ETDRe   			& .01 & - & - & - & 1385\\[.05cm]
SpecRe   			& - & - & - & 94 & 1307\\[.05cm]
ETDV   				& .017 & - & - & - & 747\\[.05cm]
\end{tabular}
}
        \caption*{(d) $\mu = 1.0393$}
    \end{subfigure}
    \quad
     \qquad
    \begin{subfigure}[c]{0.45\textwidth}
        \includegraphics[width=\textwidth, keepaspectratio=true]{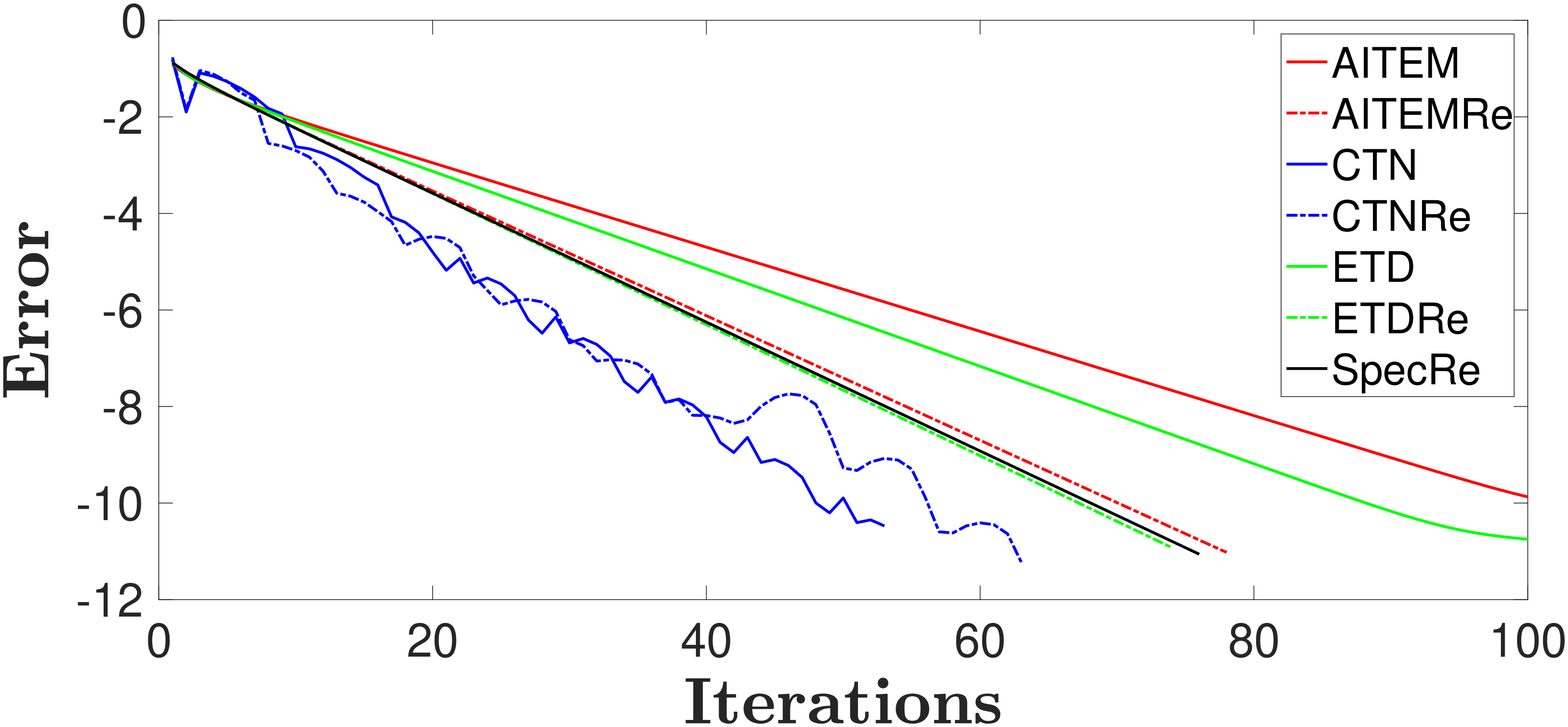}
        \caption*{(e) $V(x) =0.1x^2$ , $\sigma=1$}
    \end{subfigure}
            \quad
        \begin{subfigure}[c]{0.42\textwidth}
        {\footnotesize
        \begin{tabular}{l*{5}{c}}
Scheme              & $\Delta t$ & $c$ & $\tilde{n}$ & $r$ & Iterations\\
\hline \\[.05cm]
AITEM 				& .85 & 6 & - & - &  252 \\[.05cm]
AITEMRe          & .83 & 6 & - & -  & 78 \\[.05cm]
ACTN          		& .9 & 4 & 23 & -  & 53\\[.05cm]
ACTNRe    		& .9 & 4 & 20 & - & 63\\[.05cm]
ETD   				& .13 & - & - & - & 112\\[.05cm]
ETDRe   			& .12 & - & - & - & 74\\[.05cm]
SpecRe   			& - & - & - & 7.5 & 76 \\[.05cm]
ETDV   				& .017 & - & - & - & 695 \\[.05cm]
\end{tabular}
}
        \caption*{(f) $\mu = -1.5955$}
    \end{subfigure}
     \quad
    \begin{subfigure}[c]{0.45\textwidth}
        \includegraphics[width=\textwidth, keepaspectratio=true]{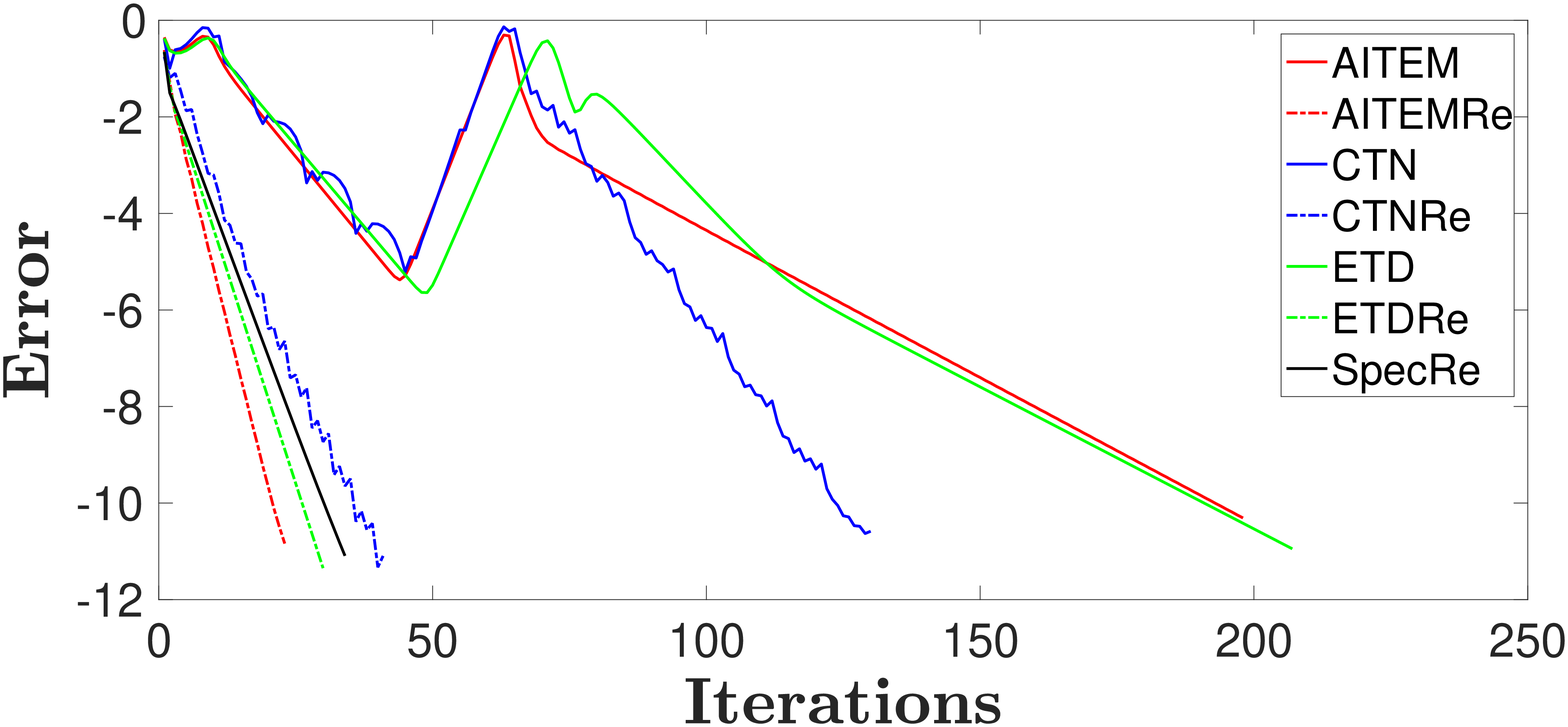}
        \caption*{(g) $V(x) =\cos(x)$ , $\sigma=1$}
         \end{subfigure}
    \quad
            \begin{subfigure}[c]{0.42\textwidth}
        {\footnotesize
        \begin{tabular}{l*{5}{c}}
Scheme              & $\Delta t$ & $c$ & $\tilde{n}$ & $r$ & Iterations\\
\hline \\[.05cm]
AITEM 				& 1.2 & 2 & - & - &  198 \\[.05cm]
AITEMRe          & 1.4 & 3 & - & -  & 23 \\[.05cm]
ACTN          		& 1.2 & 3 & 9 & -  & 130\\[.05cm]
ACTNRe    		& 1.1 & 4 & 4 & - & 41\\[.05cm]
ETD   				& .47 & - & - & - & 207\\[.05cm]
ETDRe   			& .4 & - & - & - & 30\\[.05cm]
SpecRe   			& - & - & - & 2 & 34 \\[.05cm]
ETDV   				& .016 & - & - & - & 3371 \\[.05cm]
\end{tabular}
}
        \caption*{(h) $\mu= -2.6069$ }
    \end{subfigure}
    \quad
    \begin{subfigure}[c]{0.45\textwidth}
        \includegraphics[width=\textwidth, keepaspectratio=true]{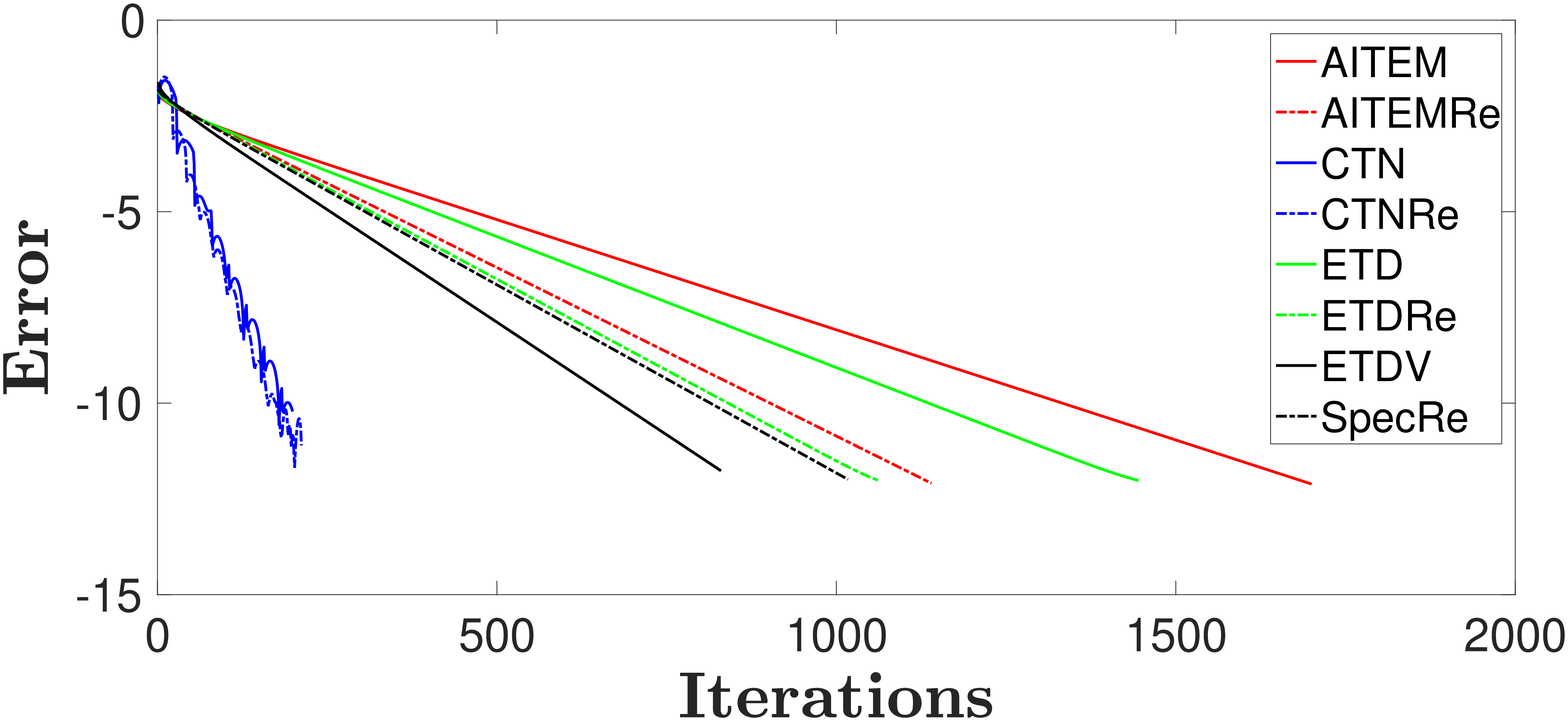}
        \caption*{(i) $V(x) = 0.01x^4 + 0.02x^2$ , $\sigma=1$}
    \end{subfigure}
    \quad
            \begin{subfigure}[c]{0.42\textwidth}
        {\footnotesize
        \begin{tabular}{l*{5}{c}}
Scheme              & $\Delta t$ & $c$ & $\tilde{n}$ & $r$ & Iterations\\
\hline \\[.05cm]
AITEM 				& .08 & 7 & - & - &  1700 \\[.05cm]
AITEMRe          & .09 & 8 & - & -  & 1140 \\[.05cm]
ACTN          		& .3 & 5 & 26 & -  & 199 \\[.05cm]
ACTNRe    		& .32 & 5 & 20 & - & 212 \\[.05cm]
ETD   				& .01 & - & - & - & 1445 \\[.05cm]
ETDRe   			& .4 & - & - & - & 1062 \\[.05cm]
SpecRe   			& - & - & - & 96 & 1017 \\[.05cm]
ETDV   				& .017 & - & - & - & 830 \\[.05cm]
\end{tabular}
}
        \caption*{(j) $\mu = -1.5795$}
    \end{subfigure}
    
            \caption{Evolution of the error, defined as the
              $L^2$ norm of the difference between successive iterates,
              as a function of the iteration index for different potentials
              $V(x)$, when seeking the ground state of the 1D NLS equation.
              The right set of tables indicates the values of the parameters
              selected and the corresponding number of iterations needed
              to reach the prescribed tolerance of $10^{-10}$.}\label{1D_1}
\end{figure}

\begin{figure}[]   
\centering
    \begin{subfigure}[c]{0.45\textwidth}
        \includegraphics[width=\textwidth, keepaspectratio=true]{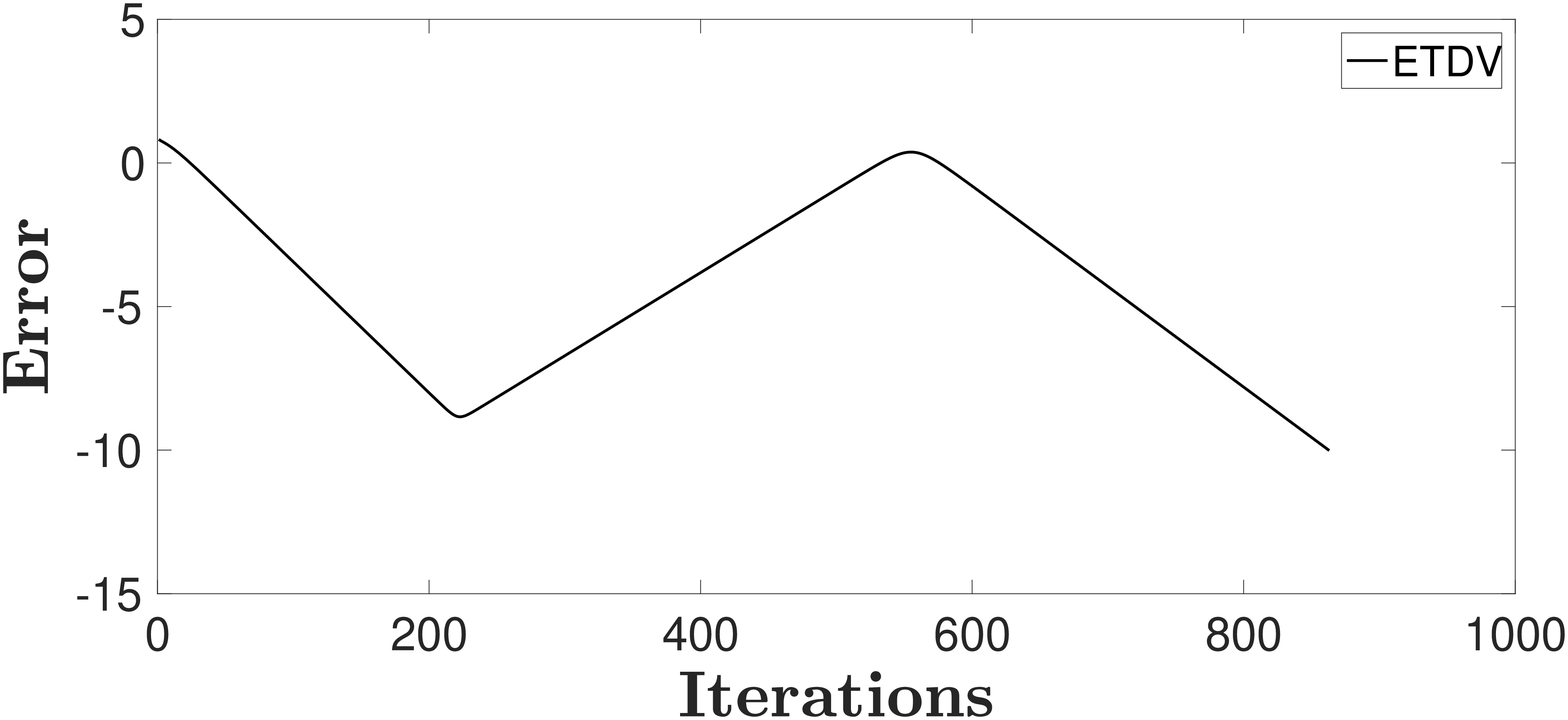}
        \caption*{(a) $V=10x^4-20x^2$, $\sigma=1$}
    \end{subfigure}
       \quad
            \begin{subfigure}[c]{0.45\textwidth}
        {\footnotesize
        \begin{tabular}{l*{5}{c}}
Scheme              & $\Delta t$ & $c$ & $\tilde{n}$ & $r$ & Iterations\\
\hline \\[.05cm]
AITEM 				& - & - & - & - &  DNC \\[.05cm]
AITEMRe          & - & - & - & - & DNC \\[.05cm]
ACTN          		& - & - & - & - & DNC \\[.05cm]
ACTNRe    		& - & - & - & - & DNC \\[.05cm]
ETD   				& - & - & - & - & DNC \\[.05cm]
ETDRe   			& - & - & - & - & DNC \\[.05cm]
SpecRe   			& - & - & - & - & DNC \\[.05cm]
ETDV   				& .016 & - & - & - & 863 \\[.05cm]
\end{tabular}
}
        \caption*{(b)}
    \end{subfigure}
            \caption{Similar to Fig.~\ref{1D_1}, but now for a double
              well potential. Only the ETDV is able to converge to
              the asymmetric ground state of this potential.} \label{1D_2}
\end{figure}

Fig.~\ref{1D_1} and \ref{1D_2} show the results of applying the methods to the cubic NLS equation 
\[\nabla^2 \psi -V(x)\psi +\sigma |\psi|^2\psi + \mu \psi=0;\]
each example corresponds to a different $V(x)$ and  $\sigma$.
Notice that examples are shown both for the focusing
case of $\sigma=1$ and for the defocusing one of $\sigma=-1$.
The diagrams on the left constitute plots of the log of the $L^2$ norm of the difference between $\psi_{n+1}$ and $\psi_n$ versus the number of iterations. We stopped all runs once the \textit{residual} error reached $10^{-10}$. The diagrams on the right show the various parameter values we used for each method as well as the total number of iterations; if a method didn't reach the prescribed tolerance, then it is labeled ``DNC'' for did not converge. 
To be precise, we do not claim that the method can not converge but rather, for the various parameter values we tried, we were not able to observe convergence. 
We also want to emphasize that although we tried to choose the parameters so that all schemes perform at their ``best", and although our results represent
the principal trend for the parameter sets examined, we cannot guarantee
that these comparisons will be valid for all possible parameter sets. Lastly, ETDV performs so poorly in some examples compared to the other 
methods that we do not always include it in the error diagrams; its total number of iterations can still be found in the relevant tables.

The general behavior shown in Fig.~\ref{1D_1} is that the continuous Nesterov methods tend to outperform the others, although AITEMRe clearly converges much quicker than the other methods in Fig.~\ref{1D_1}(g). It's also clear that the continuous Nesterov methods tend to converge quickest in the quartic potentials; this isn't surprising as CTN was devised to outperform gradient descent in poorly conditioned problems. Regardless, even for the
parabolic and periodic potentials where the iteration counts are much lower, ACTN and ACTNRe still seem to have an advantage.

ETD and ETDRe seem to perform as well as the AITEM and AITEMRe. Based only on these examples, it is not clear to us that there is a systematic advantage in using 
one method over the other. However, as we stated above, our interest in exponential time differencing is that it is an alternative way of
performing the time-stepping. 

Fig.~\ref{1D_2}, in particular, shows the possible value of 
schemes such as ETDV, as it is the only method which converges.
Overall, once again, ETD methods simply offer an efficient, alternative method of performing the time integration step.

There also does not appear to be any particular trend between
the performance of a scheme and of its renormalized version; either one can outperform the other. That being said, Fig.~\ref{1D_1}(g) is particularly interesting. All of the renormalized methods converge to an unstable state centered at the origin --where the initial
guess was also centered--.
Nevertheless, the other methods converge to the stable, ground state,
centered around $x=\pi$ i.e. around the minimum of the potential. Interestingly,
notice how this ``shift'' takes place: while initially the method
attempts to extremize by maintaining the waveform centered at the maximum,
eventually, it cannot decrease the error below a certain threshold,
being forced to seek a lower energy state by shifting the center of the
coherent structure around $x=\pi$ (see the relevant trend after
the 50th iteration), eventually decreasing the error in this new location
below the desired tolerance.

\begin{figure}[H]
\centering
    \begin{subfigure}[c]{0.45\textwidth}
        \includegraphics[width=\textwidth, keepaspectratio=true]{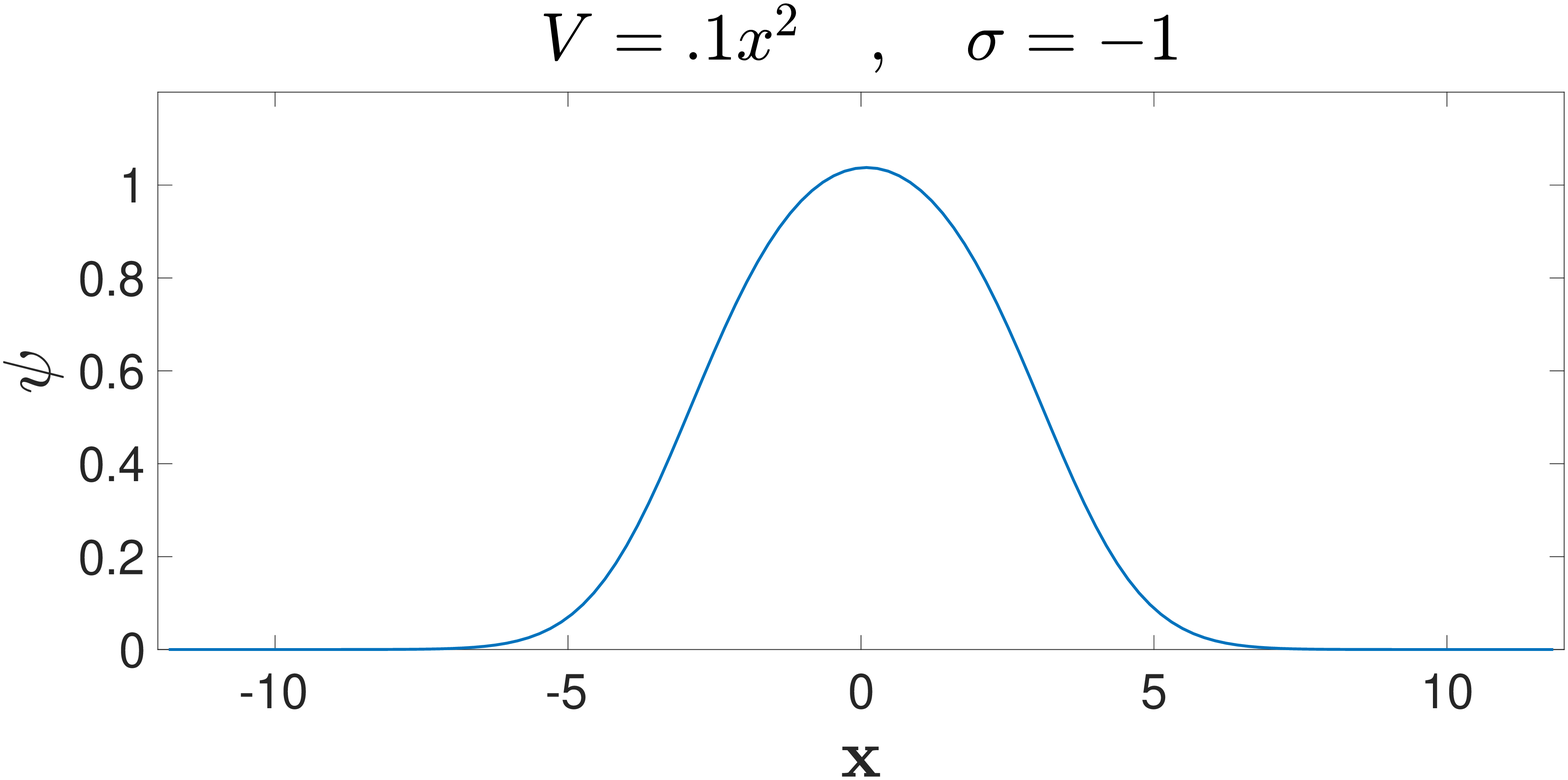}
        \caption*{(a) }
    \end{subfigure}
    \qquad
    \begin{subfigure}[c]{0.45\textwidth}
        \includegraphics[width=\textwidth, keepaspectratio=true]{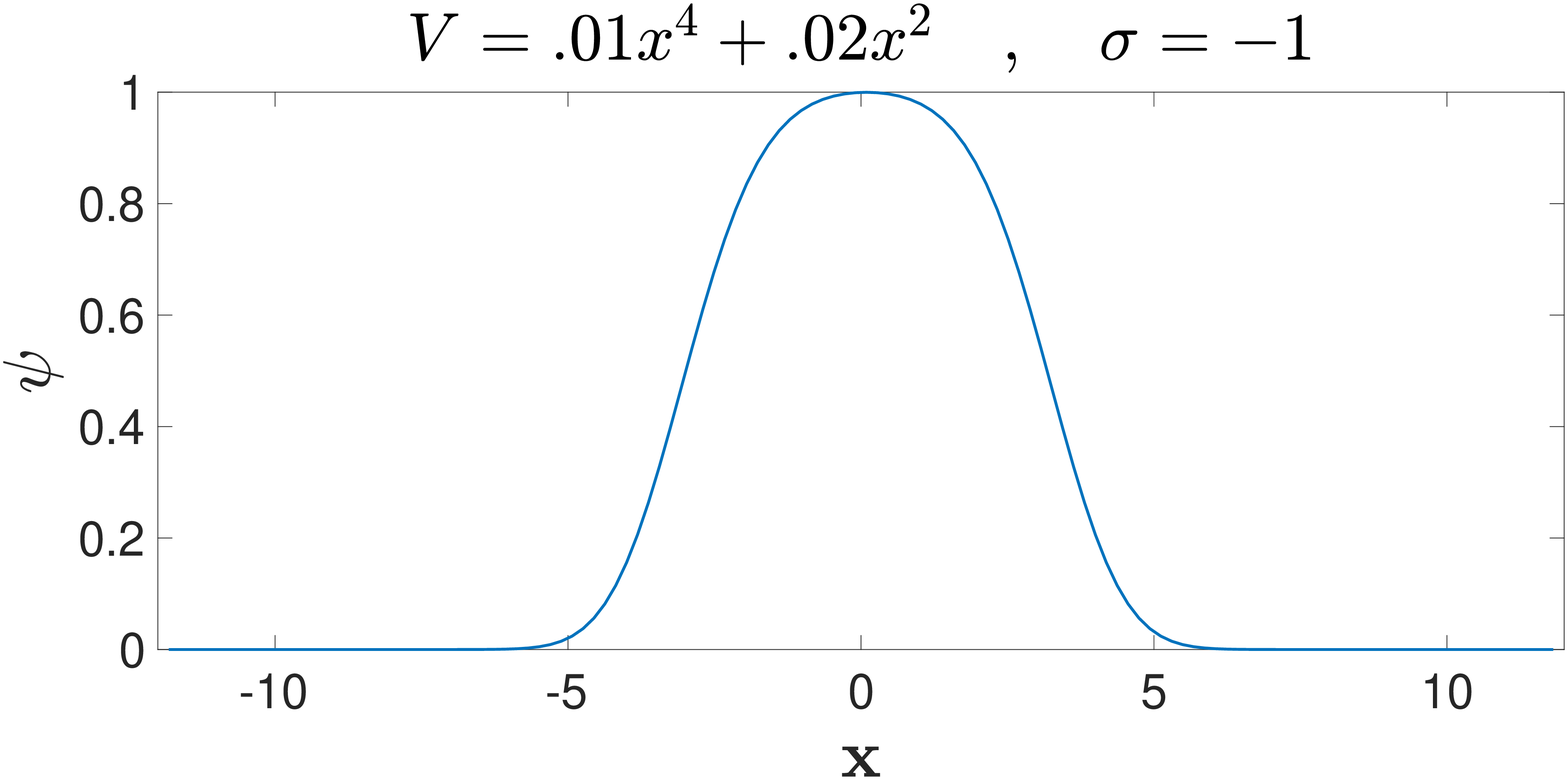}
        \caption*{(b) }
    \end{subfigure}
     \qquad
    \begin{subfigure}[c]{0.45\textwidth}
        \includegraphics[width=\textwidth, keepaspectratio=true]{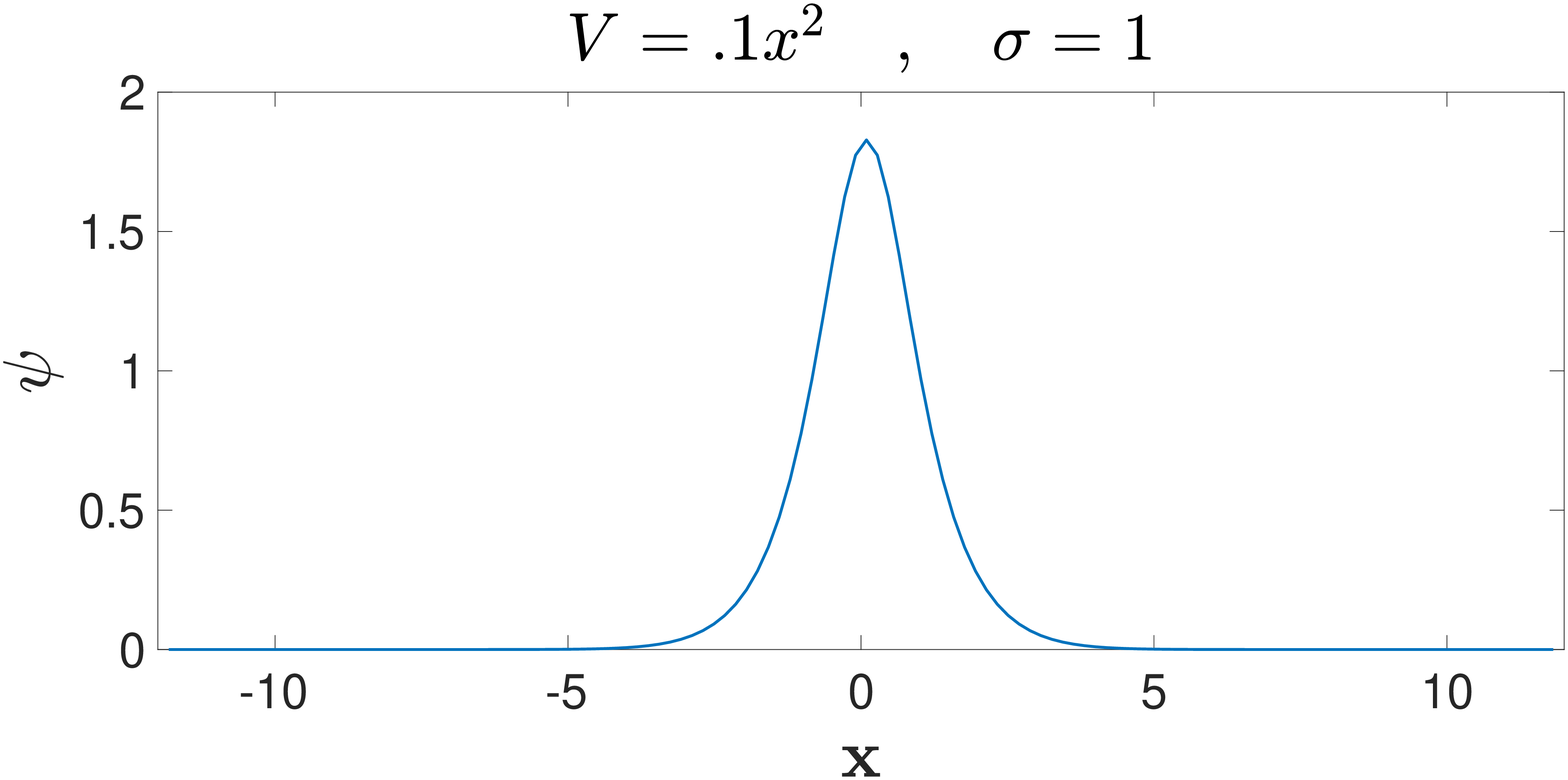}
        \caption*{(c) } 
    \end{subfigure}
     \qquad
    \begin{subfigure}[c]{0.45\textwidth}
        \includegraphics[width=\textwidth, keepaspectratio=true]{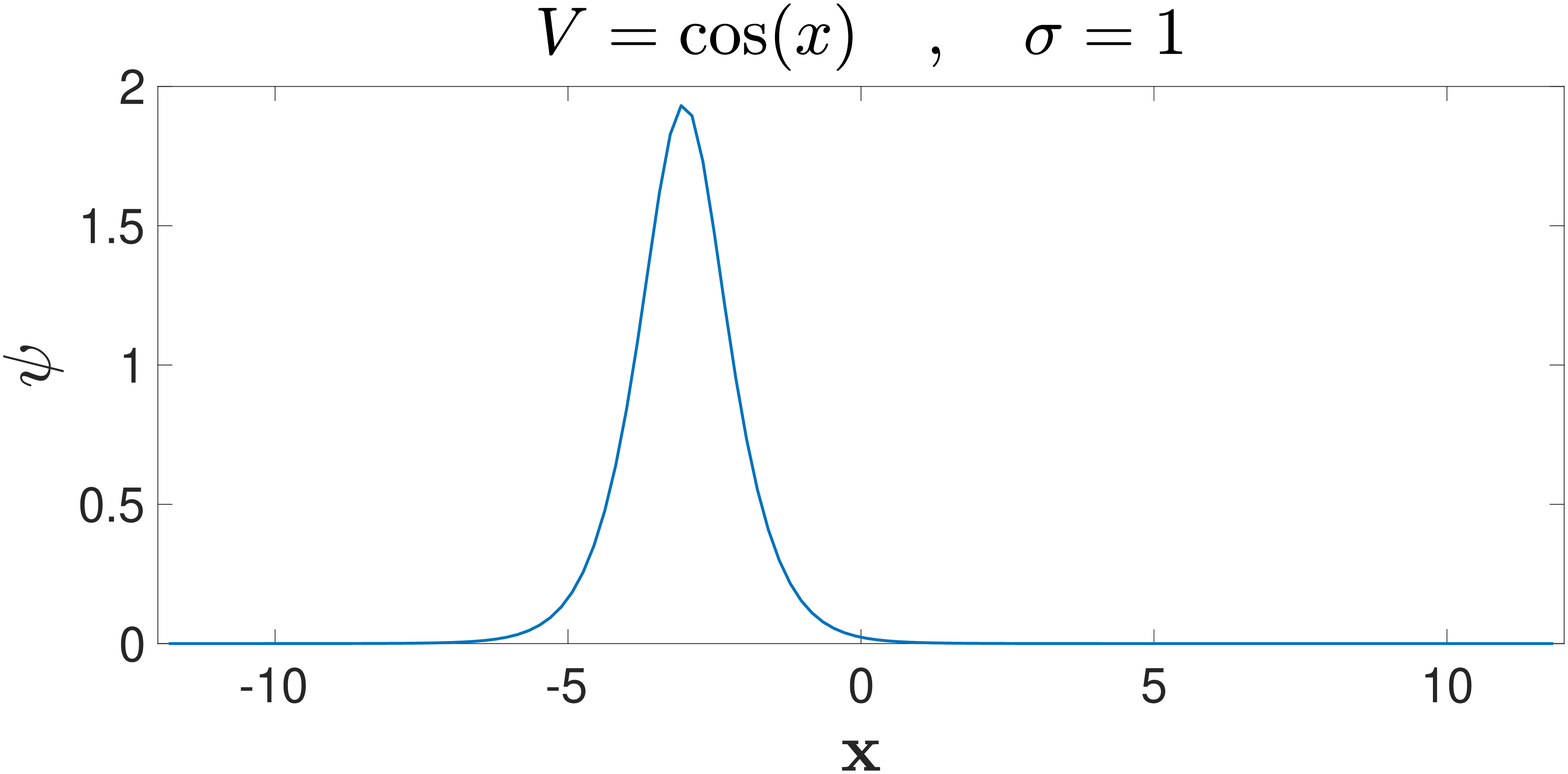}
        \caption*{(d) }
    \end{subfigure}
     \qquad
    \begin{subfigure}[c]{0.45\textwidth}
        \includegraphics[width=\textwidth, keepaspectratio=true]{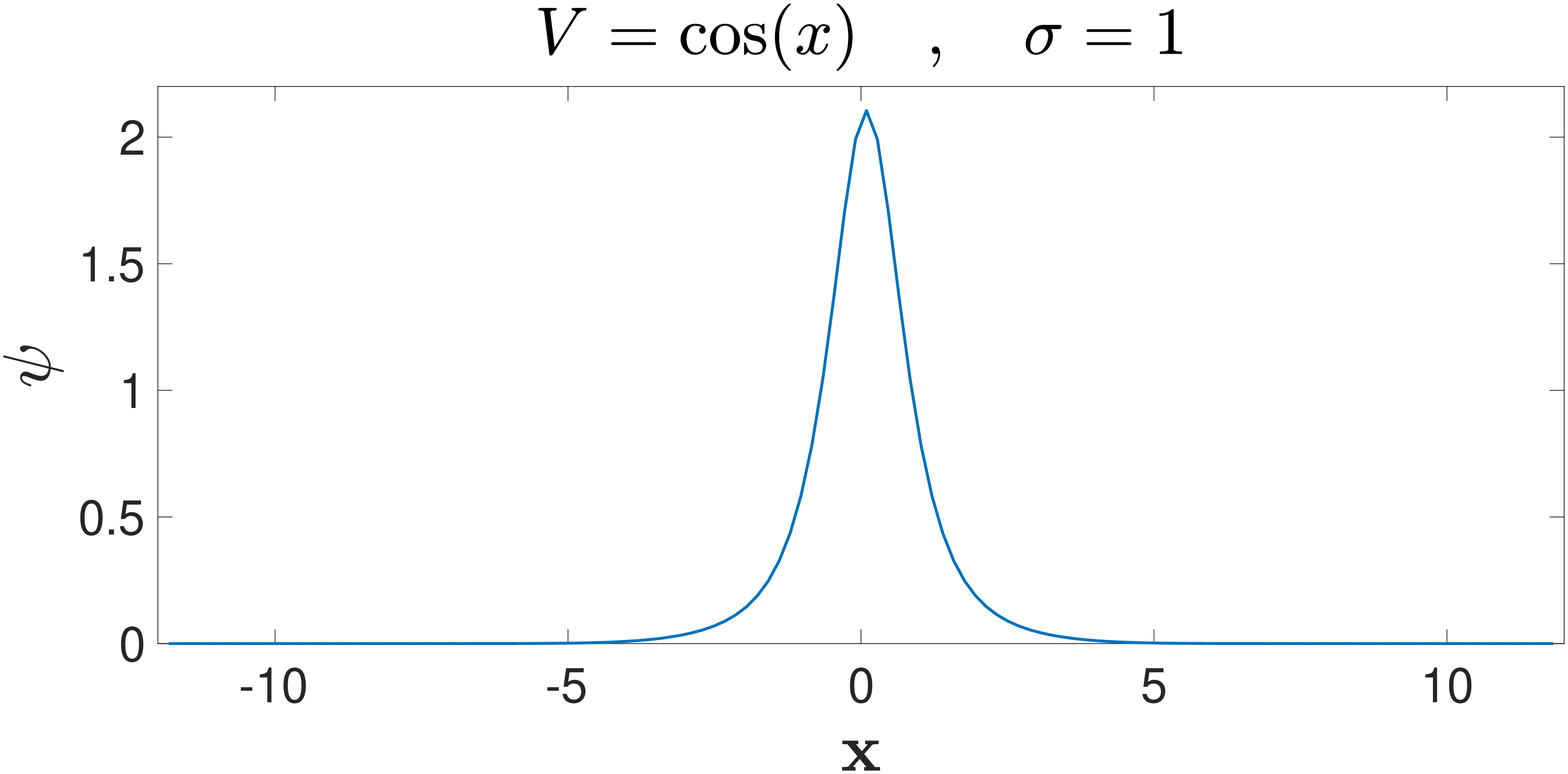}
        \caption*{(e) }
    \end{subfigure}
     \qquad
    \begin{subfigure}[c]{0.45\textwidth}
        \includegraphics[width=\textwidth, keepaspectratio=true]{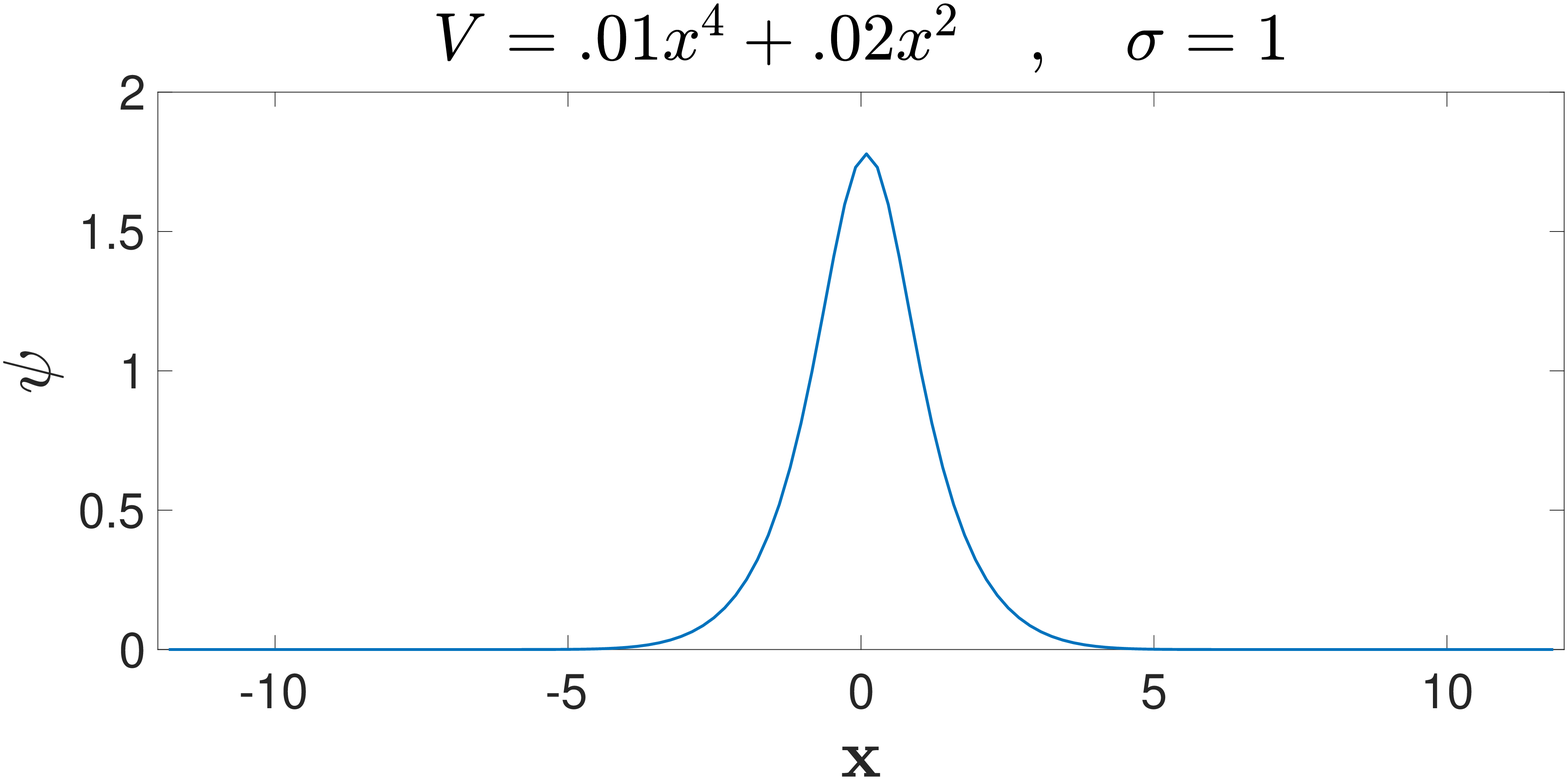}
        \caption*{(f) }
    \end{subfigure}
    \qquad
        \begin{subfigure}[c]{0.45\textwidth}
        \includegraphics[width=\textwidth, keepaspectratio=true]{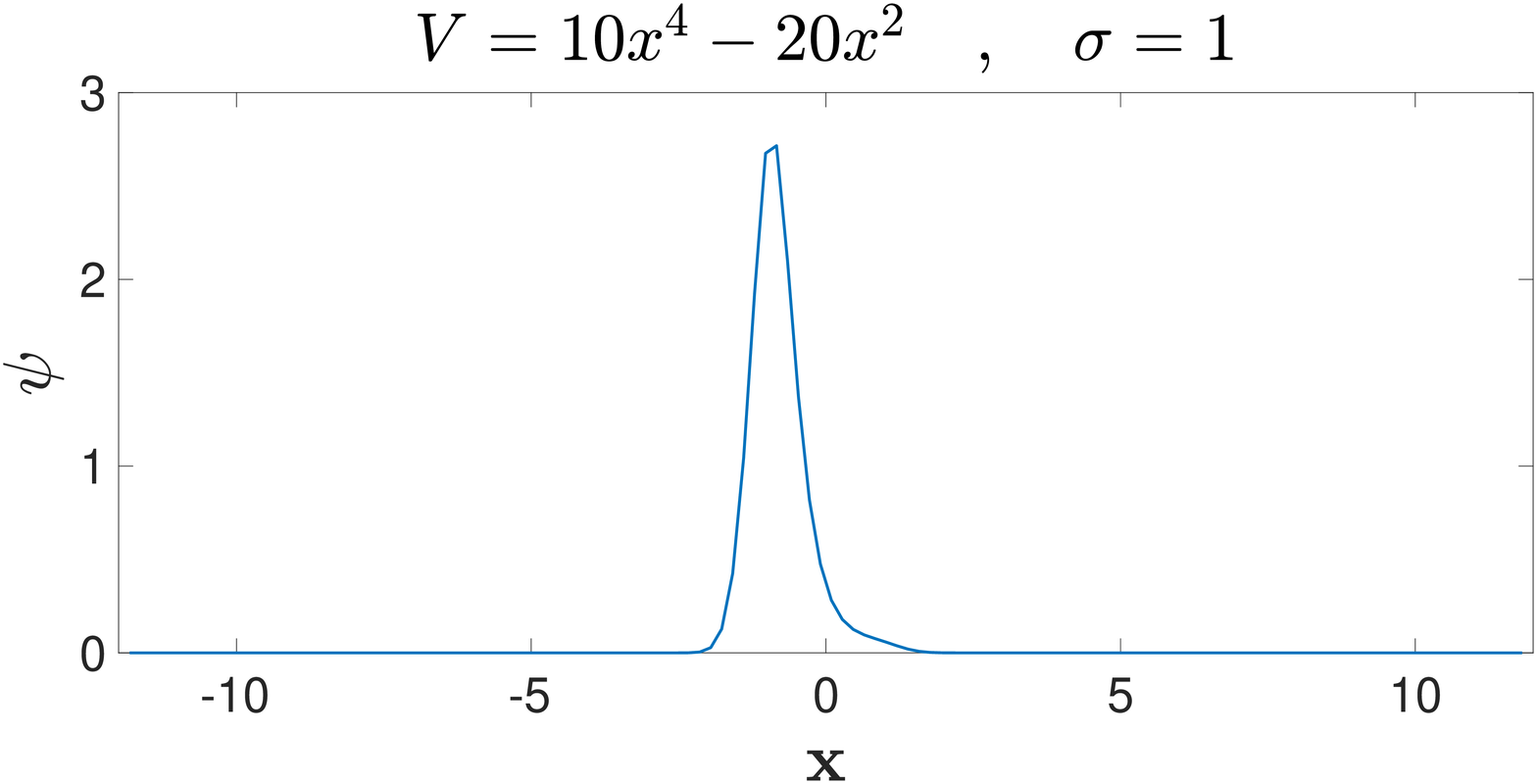}
        \caption*{(g) }
    \end{subfigure}
        \caption{Corresponding steady states of the potentials
          analyzed in the previous two figures. Notice the $x \neq 0$
        centering of cases (d) and (g).}
        \label{1d3}
\end{figure}

The case reported in Fig.~\ref{1D_2} bears some similarities to the above
described scenario, as once again the state is initialized as located
at the center, yet the double well nature of the potential does not
favor such a localization at the maximum. Instead, the lowest energy
state consists of a concentration of the atoms (or the optical power)
in either the left or right well of the relevant potential.
This symmetry-breaking is a feature well-known
in the context of double-well potentials~\cite{rcg:BEC_book2}.
The ETDV
attempts for a while to extremize the free energy via localization
at the center. Eventually, being unsuccessful, it is led to shift
the wave mass to one of the two sides converging to the state shown
in panel (g) of Fig.~\ref{1d3}. This figure contains the ground state
identified in all the cases of Figs.~\ref{1D_1}-\ref{1D_2}, rendering
transparent that in case (d) and (g), the localization happens around
$x \neq 0$.

\subsection{Ground States in 2D}
\begin{figure}[]
\centering
    \begin{subfigure}[c]{0.45\textwidth}
        \includegraphics[width=\textwidth, keepaspectratio=true]{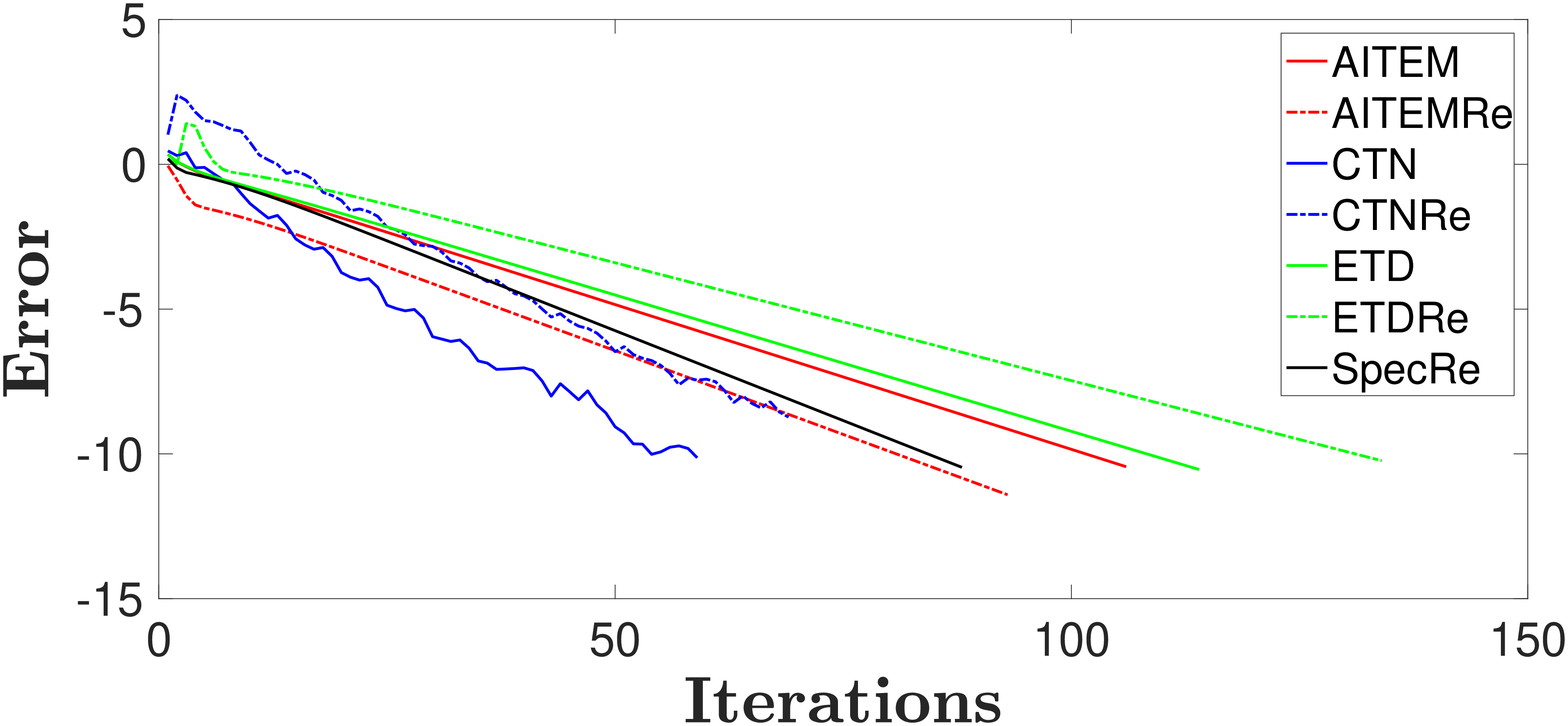}
        \caption*{(a)}
    \end{subfigure}
                \begin{subfigure}[c]{0.45\textwidth}
        {\footnotesize
        \begin{tabular}{l*{5}{c}}
Scheme              & $\Delta t$ & $c$ & $\tilde{n}$ & $r$ & Iterations\\
\hline \\[.05cm]
AITEM 				& 1 & 3 & - & - &  106 \\[.05cm]
AITEMRe          & 1.7 & 4 & - & - & 93 \\[.05cm]
ACTN          		& 1.1 & 2 & 5 & - & 59 \\[.05cm]
ACTNRe    		& 1.1 & 2 & 7 & - & 69 \\[.05cm]
ETD   				& .3 & - & - & - & 114 \\[.05cm]
ETDRe   			& .3 & - & - & - & 134 \\[.05cm]
SpecRe   			& - & - & - & 2.1& 88 \\[.05cm]
\end{tabular}
}
        \caption*{(b)}
    \end{subfigure}
    \quad
        \begin{subfigure}[c]{0.45\textwidth}
        \includegraphics[width=\textwidth, keepaspectratio=true]{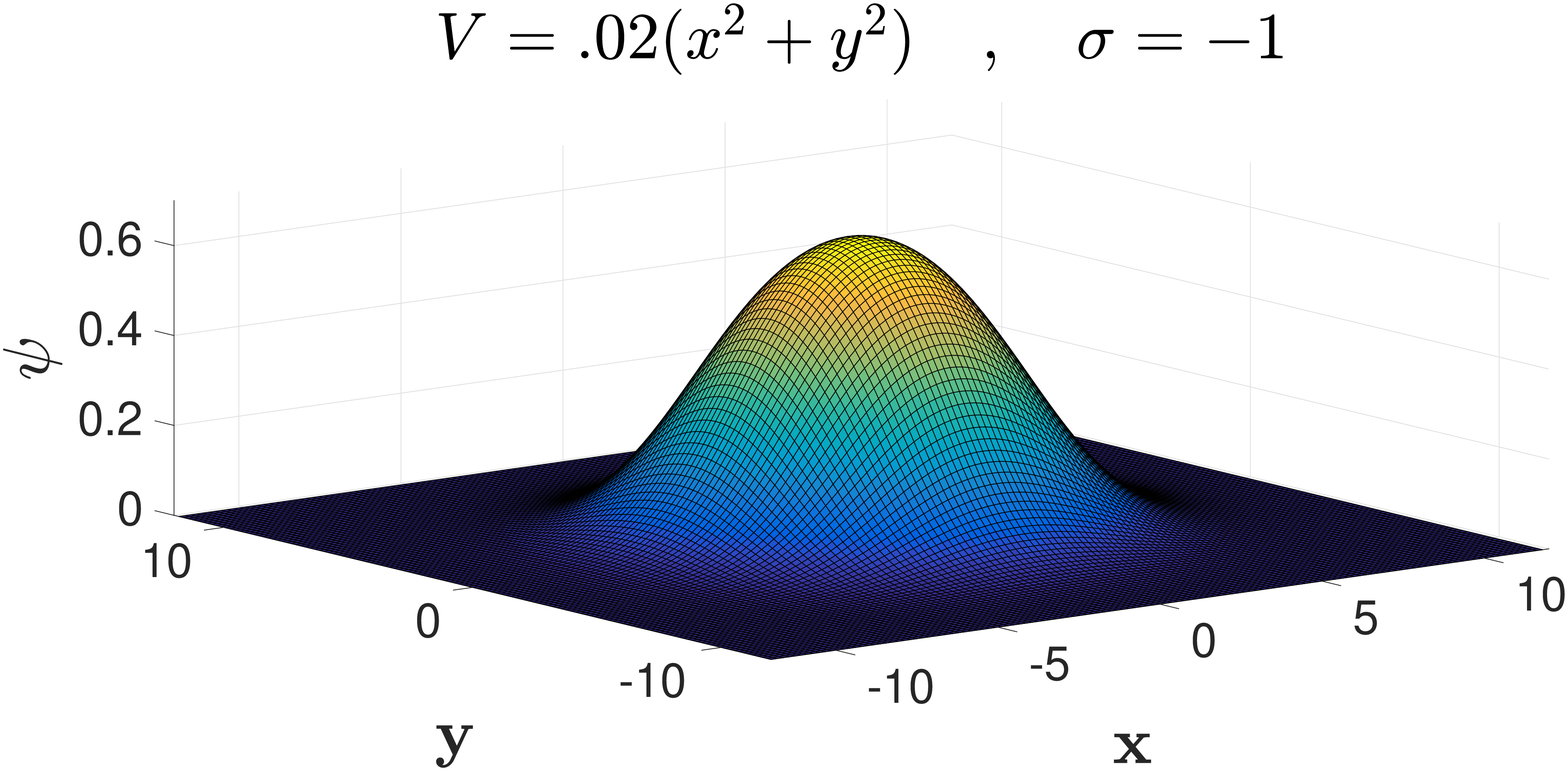}
        \caption*{(c)}
    \end{subfigure}

    \begin{subfigure}[c]{0.45\textwidth}
        \includegraphics[width=\textwidth, keepaspectratio=true]{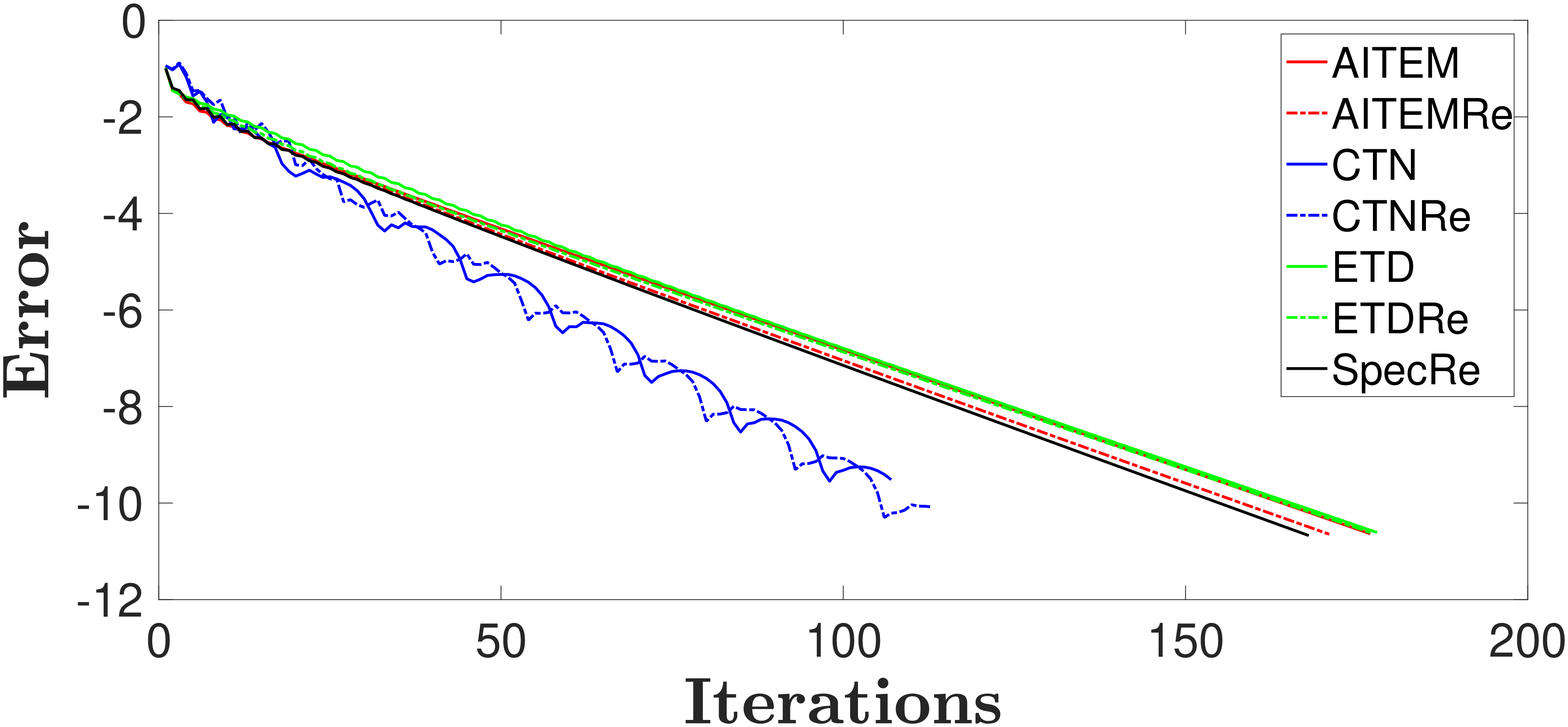}
        \caption*{(d)}
    \end{subfigure}
    \quad
     \begin{subfigure}[c]{0.45\textwidth}
        {\footnotesize
        \begin{tabular}{l*{5}{c}}
Scheme              & $\Delta t$ & $c$ & $\tilde{n}$ & $r$ & Iterations\\
\hline \\[.05cm]
AITEM 				& 1.1 & 3 & - & - &  177 \\[.05cm]
AITEMRe          & 1.1 & 3 & - & - & 171 \\[.05cm]
ACTN          		& 1 & 2 & 5 & - & 107 \\[.05cm]
ACTNRe    		& 1 & 2 & 7 & - & 113 \\[.05cm]
ETD   				& .31 & - & - & - & 178 \\[.05cm]
ETDRe   			& .3 & - & - & - & 177 \\[.05cm]
SpecRe   			& - & - & - & 2.5 & 168 \\[.05cm]
\end{tabular}
}
        \caption*{(e)}
    \end{subfigure}
    \quad
        \begin{subfigure}[c]{0.45\textwidth}
        \includegraphics[width=\textwidth, keepaspectratio=true]{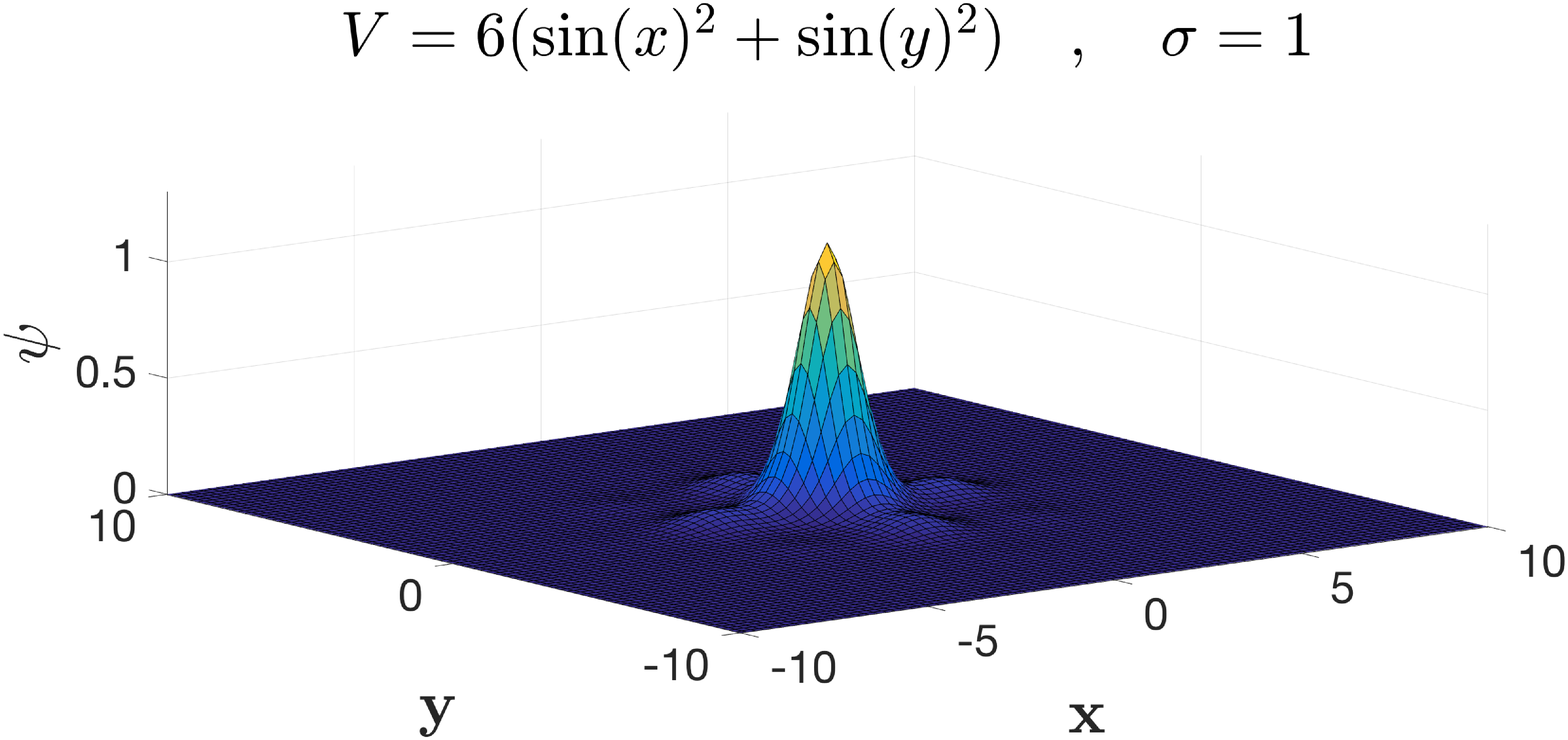}
        \caption*{(f)}
    \end{subfigure}
        \caption{Two prototypical case examples in 2D. The top set
          of panels (a)-(c) displays the evolution of the error
          over the number of iterations, the parameters (and
          convergence iteration number) of the different methods,
          and the profile of the resulting solution for a parabolic
          trap in a defocusing 2D NLS
          with a Gaussian initial guess. Panels (d)-(f) report in similar format but now
        for a focusing 2D NLS with a periodic potential.}\label{2D_1}
\end{figure}

In this section we focus on the 2D variant of the NLS equation, once again
attempting to identify the ground state of the nonlinear elliptic problem.
Fig.~\ref{2D_1}(a,b,c) is a defocusing NLS equation with quadratic potential. For the initial condition we use $\psi_0= A e^{-(x^2 + y^2)}$, where $A$ is chosen so that the resulting power is $P=17$.
Here, the ground single-hump state (whose linear limit is proportional to
the initial guess) is rapidly converged upon.
Fig.~\ref{2D_1}(d,e,f) is a focusing NLS equation with periodic potential and we use a similar initial condition except $A$ was chosen so that the chemical potential is $\mu=3.7$. In this case, all the schemes converge
in a comparable number of iterations to a gap soliton solution of
the problem.

As in the 1D case, the same general trends tend to hold. The continuous Nesterov methods seem to outperform the others, the ITEM schemes and ETD schemes 
seem to not have significant differences in their performance,
and again there does not seem to be definitive preferentiability
manifested between renormalized methods and their standard version. 

\subsection{Excited States in 1D}

Naturally, it is of substantial interest to go beyond the most fundamental
states and seek excited states in the system. E.g. both in the
atomic~\cite{GP,becbook1,becbook2,rcg:BEC_book2} and in the optical
problem~\cite{kivshar}, excited states such as dark solitons and
multi-solitons in 1D and vortices and related structures (such
as ring or planar dark solitons) in higher
dimensions have been of particular interest.

In this section we combine ACTN with the so-called Squared Operator Method (SOM)\citep{9} in order to capture such excited states. We quickly recap the basic idea: consider the gradient flow applied to some function $F$
\[\dot{u}=-F(u).\]
Naturally, this will only converge to local minima (in the case that $F$ is the gradient of some function) or, more generally, to a steady state having only eigenvalues with negative real part (if $F$ is not the gradient of some function). To extend this method to other steady states, one can instead consider the system
\[\dot{u}=-DF(u)F(u).\]

One quickly sees that every steady state of $F$  is a steady state of $DF(u)F(u)$ and, by taking the derivative of the RHS, one sees that every steady state of $F$ is stable in this new system. Hence, the SOM converges to every steady state of $F$ provided the initial condition is sufficiently close. Using CTN instead of the gradient flow, we get
\[\ddot{u} + \frac{3}{t}u  + DF(u)F(u) = 0.\]
It is this equation that we will study in what follows, and to which we will refer to as Squared (Operator) Continuous Time Nesterov (SCTN).

As an initial test, we seek families of stationary states of 
\[\nabla^2 \psi -0.1x^2 \psi - \psi^3 + \mu \psi = 0\]
i.e., tackling the defocusing problem with a parabolic trap,
in the spirit of earlier works such as~\cite{tristram,alfimov}.
We apply the ACTN method to the SCTN equation, resulting in the iteration
\begin{IEEEeqnarray}{rcl}\label{ASCTN}
M &=& (c - \nabla ^2)^2 \nonumber \\[.2cm]
\mu_n & \; = \; & \frac{\langle L \psi_n, \psi_n \rangle}{\langle \psi_n, \psi_n\rangle} \nonumber \\[.2cm]
\tilde{\psi}_{n+1} &=& (2- \frac{3}{\tilde{n}})\psi_n + (\Delta t)^2 M^{-1} (\nabla^2  -V(x) - 3 \psi_n^2 + \mu_n)(\nabla^2 \psi_n -V(x) \psi_n - \psi_n^3 + \mu_n \psi_n) \nonumber \\
 &&- (1-\frac{3}{\tilde{n}})\psi_{n-1} \\[.2cm]
\psi_{n+1} &\;=\;& \tilde{\psi}_{n+1} \sqrt{\frac{P}{\langle \tilde{\psi}_{n+1},\tilde{\psi}_{n+1} \rangle}} \nonumber
\end{IEEEeqnarray}
which we will refer to as ASCTN.

\begin{figure}[h]
\centering
    \begin{subfigure}[c]{0.45\textwidth}
        \includegraphics[width=\textwidth, keepaspectratio=true]{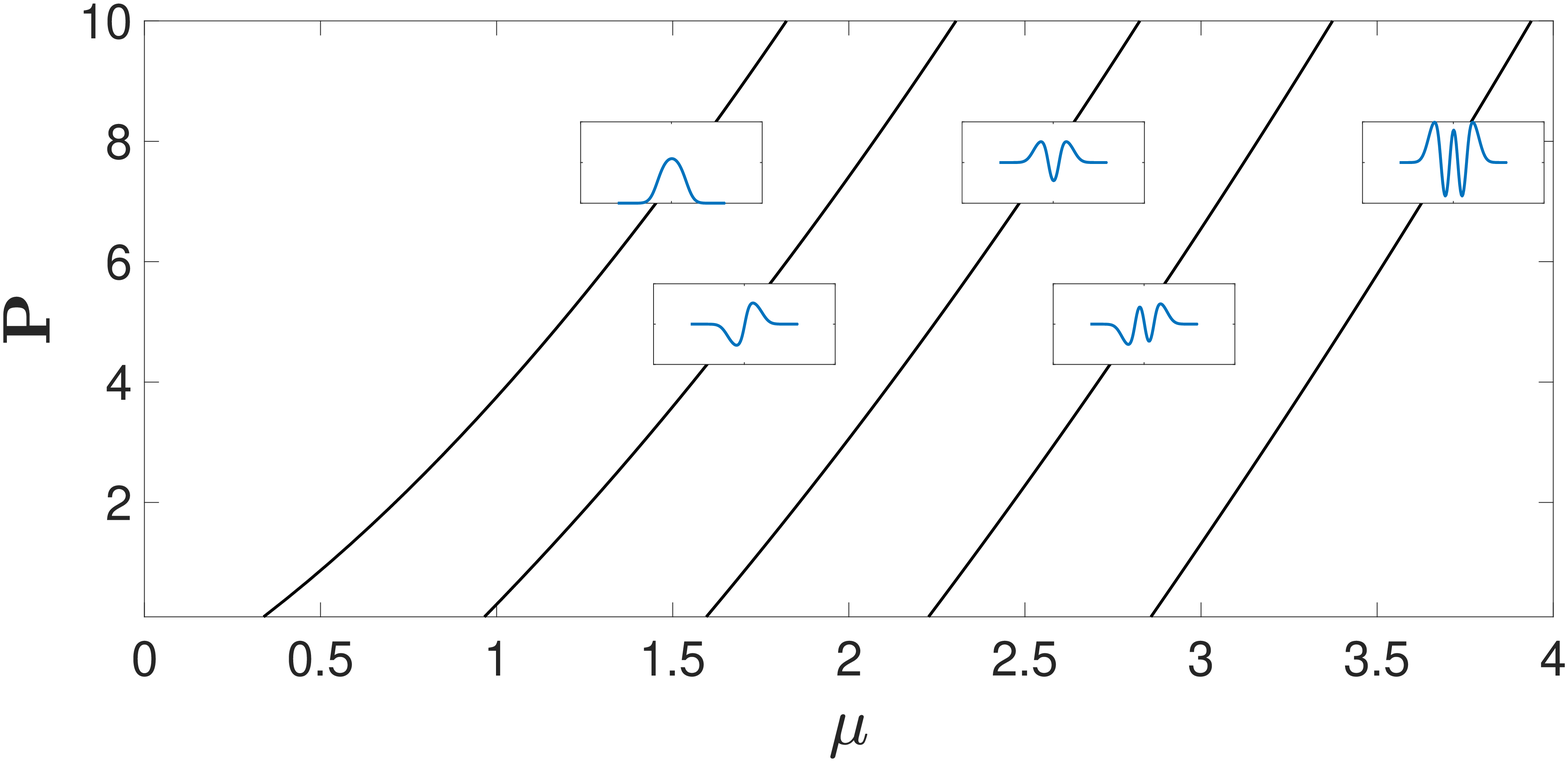}
        \caption*{(a)}
    \end{subfigure}
    \qquad
    \begin{subfigure}[c]{0.45\textwidth}
        \includegraphics[width=\textwidth, keepaspectratio=true]{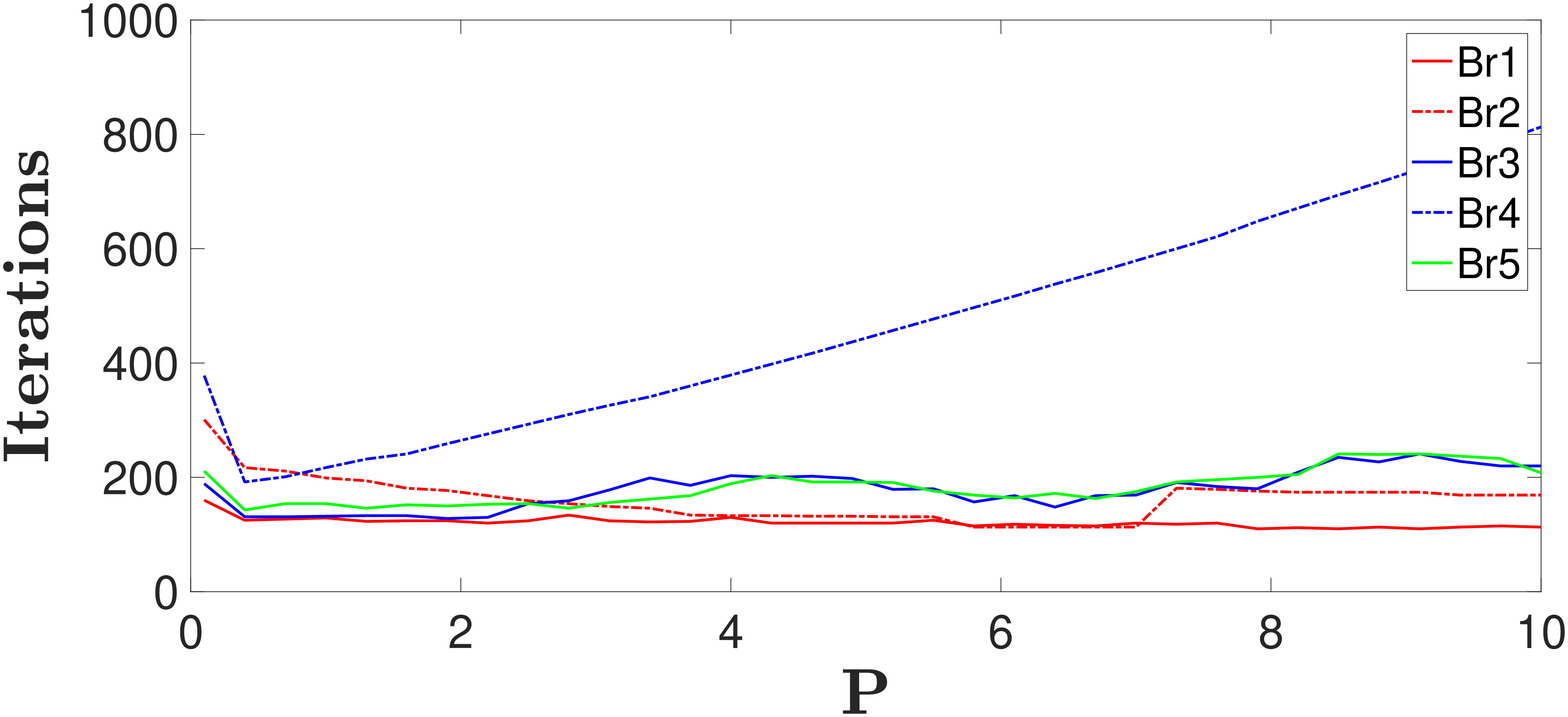}
        \caption*{(b)}
    \end{subfigure}
    \caption{Left panel: The bifurcation diagram of the first five
      excited states of the defocusing 1D NLS with parabolic trap.
      Right panel: Number of iterations needed to go from each point
    on the relevant branch to the next.}\label{1D_Excited}
\end{figure}

Fig.~\ref{1D_Excited}(a) is the aforementioned bifurcation diagram
(in a format similar to that of~\cite{alfimov}),
shown here with five branches. For each branch, we started the continuation near $P=0$ and used a combination of Gaussians as our initial guess; knowledge of the corresponding linear Schr{\"o}dinger equation's eigenfunctions would also work well, as is done in the next subsection. Once the method converges, we increase the value of $P$ by $\Delta P =0.3$ and then use the previous state as the new initial condition (in the spirit of parametric continuation). Fig.~\ref{1D_Excited}(b) shows the number of iterations necessary to go from one point on a branch to the next point on the branch (as a function of P); aside from branch 3, we see that it generally takes between 150 to 300 iterations to converge. 

We also want to mention that we performed ASCTN in Fourier space as well (similar to ACTN). Because of this, the action of the jacobian is relatively cheap to calculate and so one doesn't need to store any large matrices. On the other hand, if one were doing finite differences/elements, one could instead use \cite{CTK} the approximation
\begin{IEEEeqnarray}{rcl}\label{ACTN2}
DF(u)F(u) &=& \frac{d}{d\epsilon} \bigg[ F(u + \epsilon F(u) \bigg] \bigg |_{\epsilon=0}  \nonumber \\[.4cm]
&\approx& \frac{F(u+\epsilon F(u)) - F(u)}{\epsilon} \nonumber 
\end{IEEEeqnarray}
which again eliminates the need to form the Jacobian. This
significantly decreases the cost of the relevant numerical computation.

\subsection{Excited States in 2D}
Finally, we briefly wish to test the effectiveness of the ASCTN method in the 2D realm. Following the recent work of~\citep{10}, we study the NLS equation
\[\frac{1}{2}\nabla^2 \psi -0.02(x^2+y^2) \psi - |\psi|^2 \psi + \mu \psi = 0\]
In the limit as $P \to 0$, the nonlinearity becomes irrelevant and the stationary states bifurcate out of the linear limit. These linear eigenfunctions can be represented in the form \citep{11}
\[|m,n\rangle :=\psi_{m,n} = C H_m(\sqrt{0.2}x)H_m(\sqrt{0.2}y) e^{-0.1(x^2+y^2)}\]
where $C$ is some constant, $m,n$ are nonnegative integers, and $H_m$ is the $m$-the Hermite polynomial. We note that the corresponding value of the
linear eigenvalue $\mu$ of the corresponding states parametrized
by the quantum numbers $m$ and $n$ is given by
\[\mu_{m,n} = 0.2(m + n +1).\]

\begin{figure}
\centering
    \begin{subfigure}[c]{0.6\textwidth}
        \includegraphics[width=\textwidth, keepaspectratio=true]{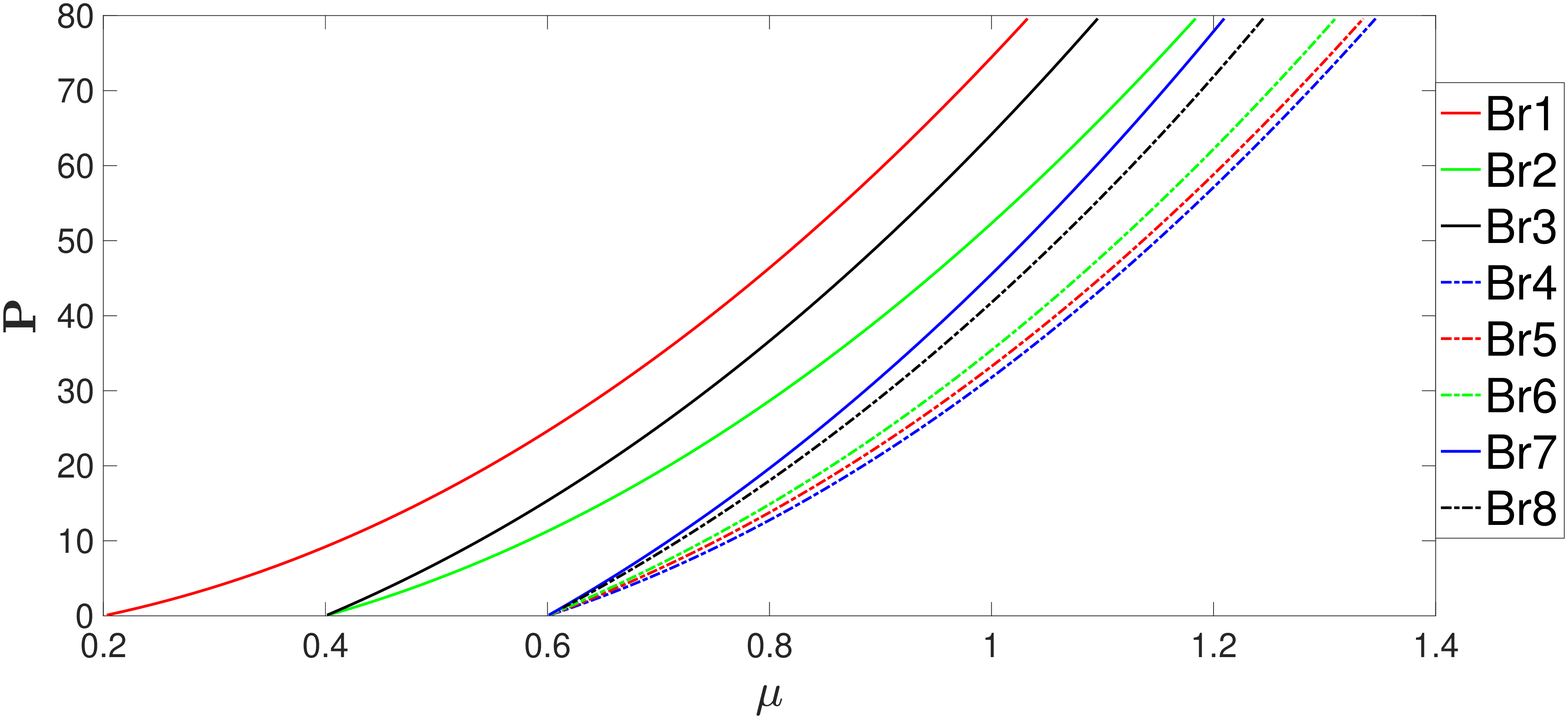}
        \caption*{(a)}
    \end{subfigure}
         \qquad
    \begin{subfigure}[c]{0.6\textwidth}
        \includegraphics[width=\textwidth, keepaspectratio=true]{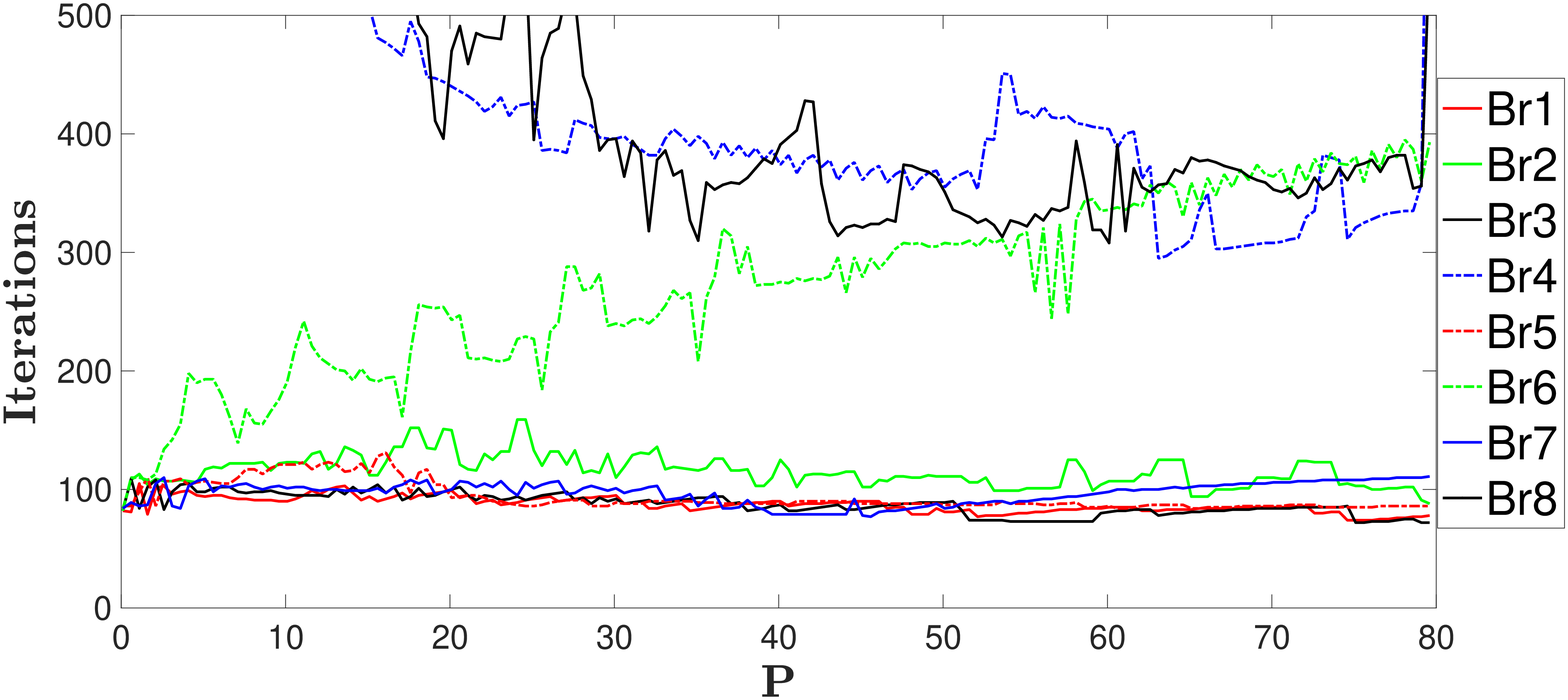}
        \caption*{(b)}
    \end{subfigure}
    \caption{Panel (a) illustrates the different branches
      identified in the two-dimensional bifurcation diagram
      of the elliptic NLS problem with the parabolic trap. The bottom panel
      shows the number of iterations needed for the ASCTN scheme
      to converge from one solution (member of a branch)
      to the next (member of the branch).} \label{fig:2D_Bif}
\end{figure}

\begin{figure}[H]
\centering
    \begin{subfigure}[c]{0.3\textwidth}
        \includegraphics[width=\textwidth, keepaspectratio=true]{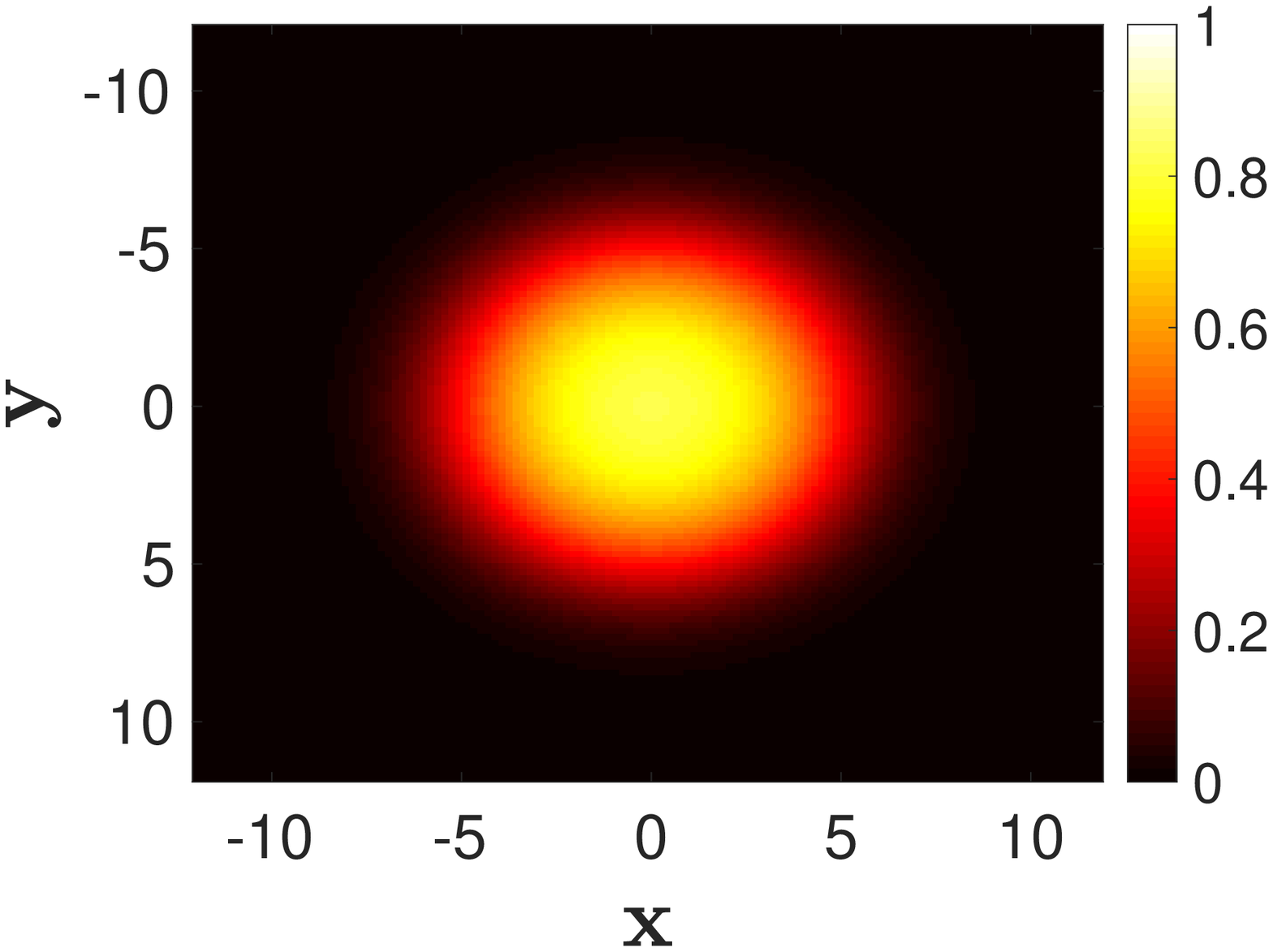}
        \caption*{(a) Br1: $|0,0\rangle$}
    \end{subfigure}
    \begin{subfigure}[c]{0.3\textwidth}
        \includegraphics[width=\textwidth, keepaspectratio=true]{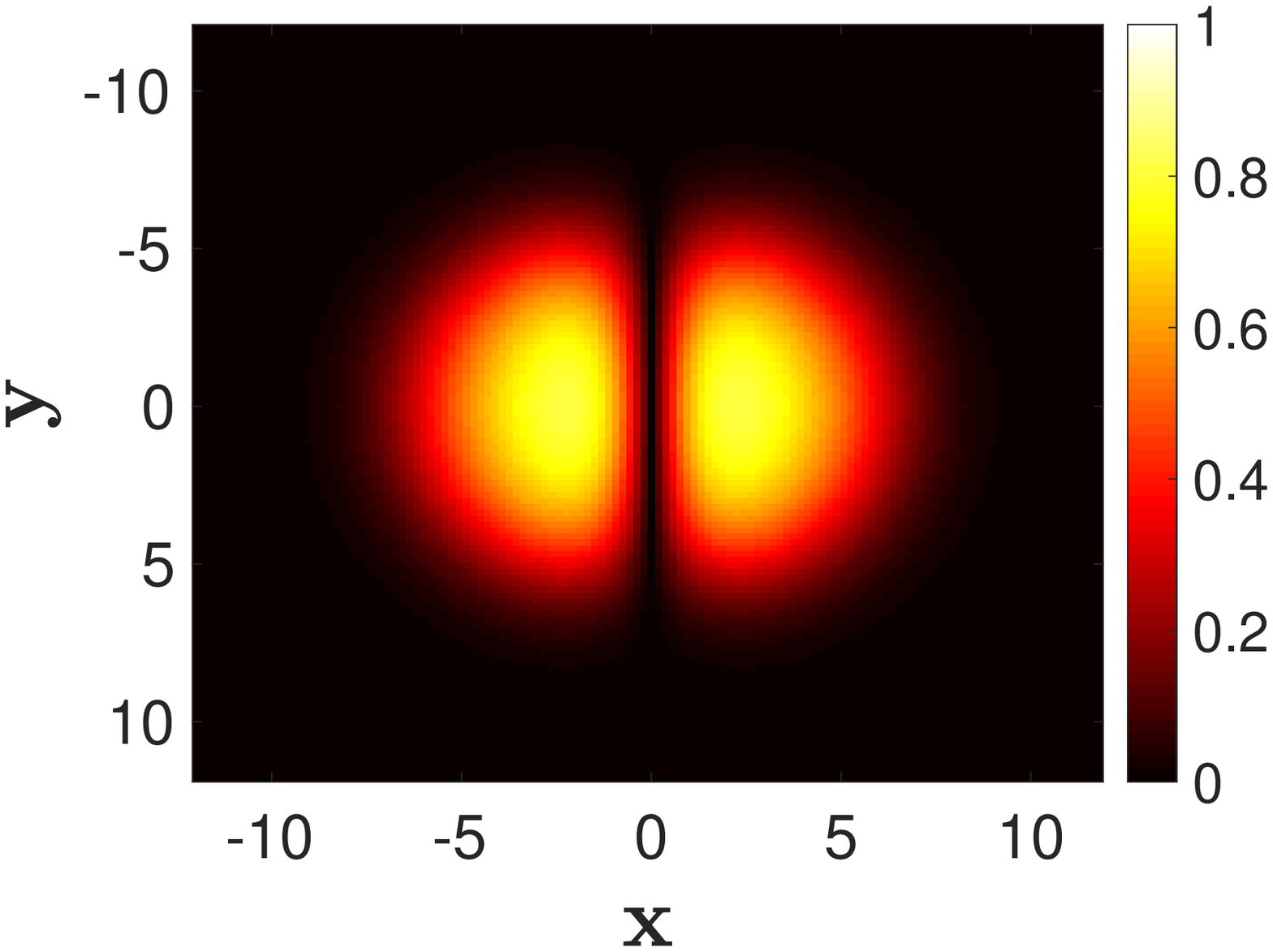}
        \caption*{(b) Br2: $|1,0\rangle$}
    \end{subfigure}
    \begin{subfigure}[c]{0.3\textwidth}
        \includegraphics[width=\textwidth, keepaspectratio=true]{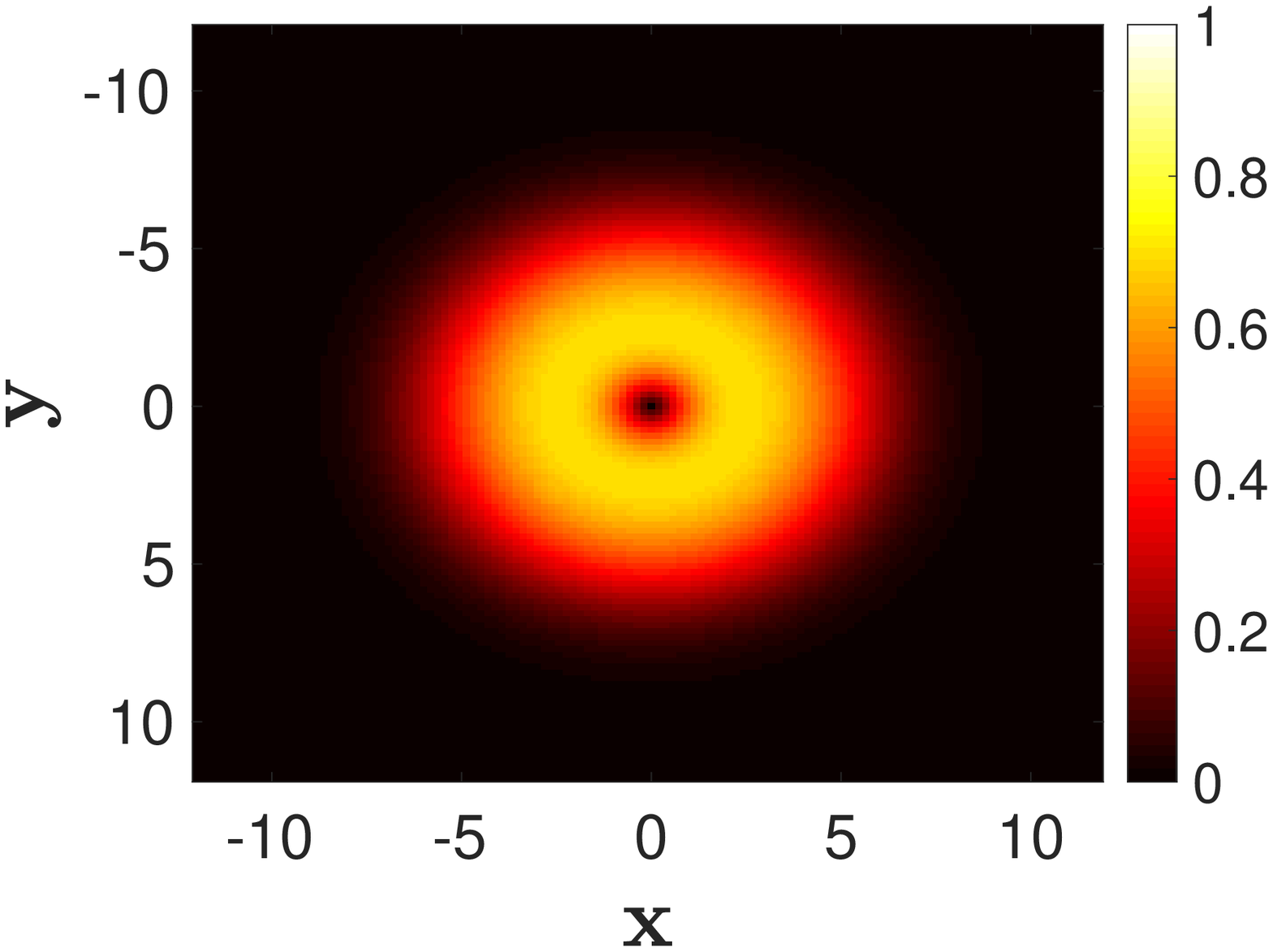}
        \caption*{(c) Br3:  $|1,0\rangle + i |0,1\rangle$} 
    \end{subfigure}
    \begin{subfigure}[c]{0.3\textwidth}
        \includegraphics[width=\textwidth, keepaspectratio=true]{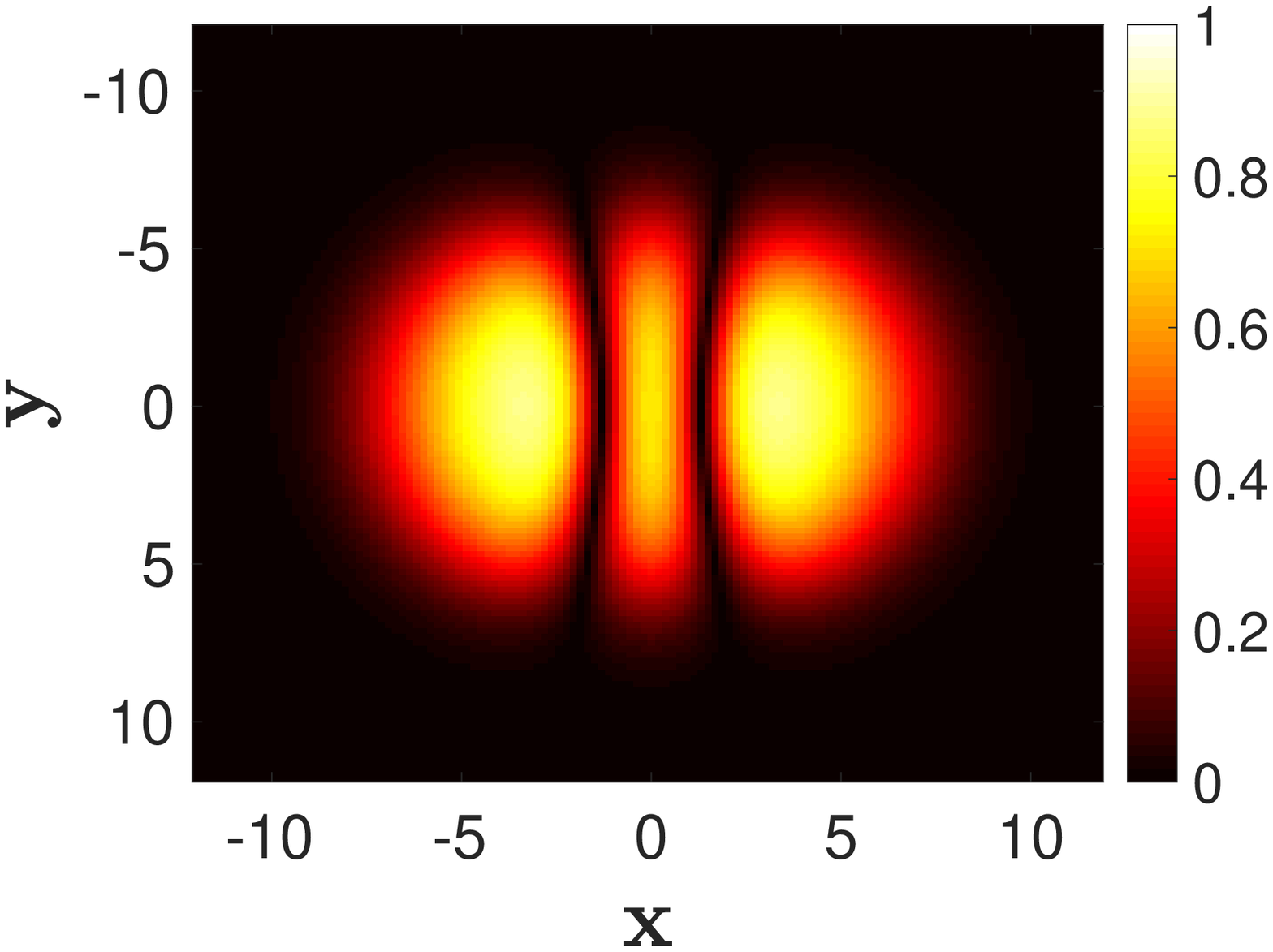}
        \caption*{(d) Br4:  $|2,0\rangle$}
    \end{subfigure}
    \begin{subfigure}[c]{0.3\textwidth}
        \includegraphics[width=\textwidth, keepaspectratio=true]{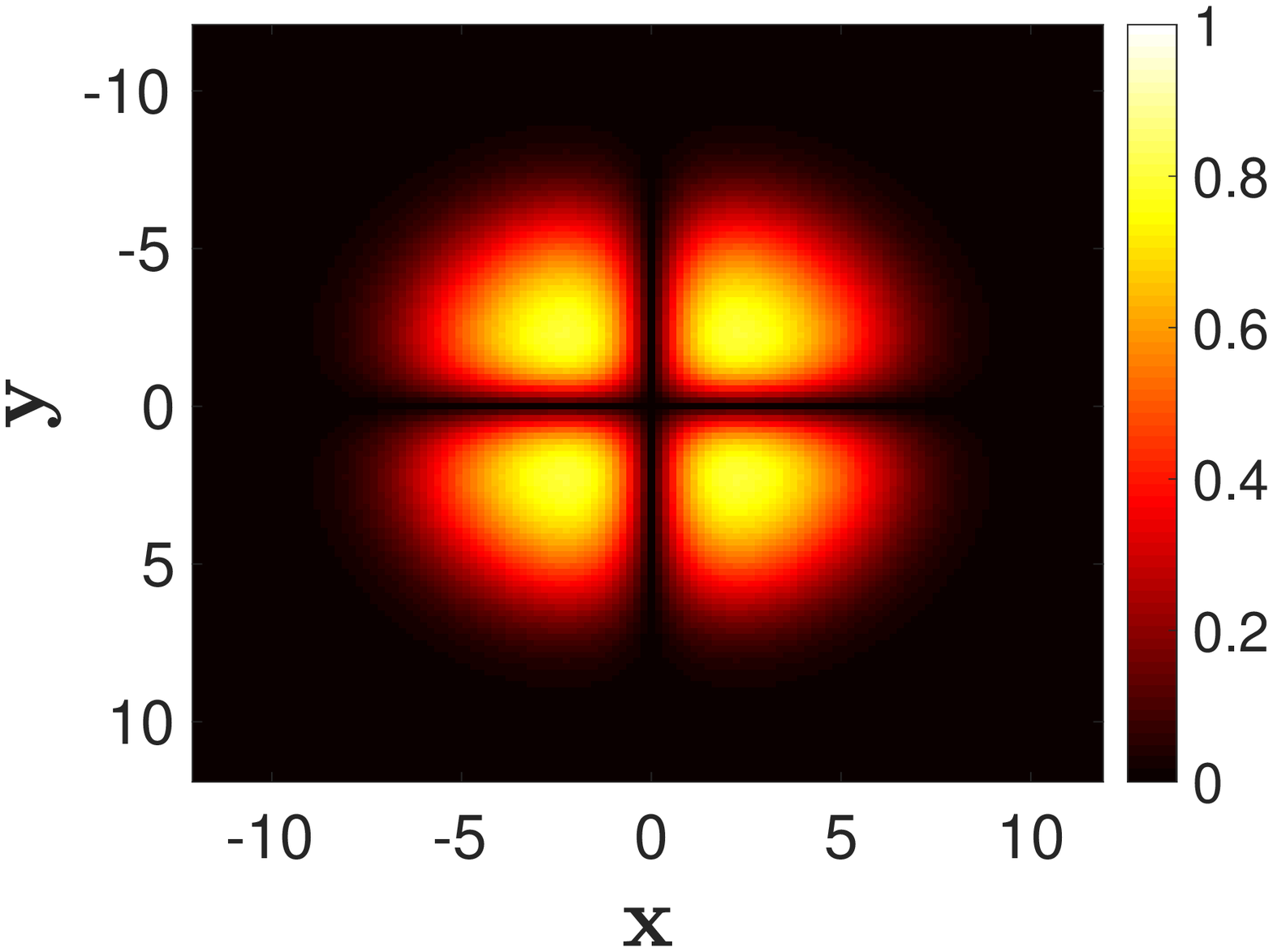}
        \caption*{(e) Br5:  $|1,1\rangle$}
    \end{subfigure}
    \begin{subfigure}[c]{0.3\textwidth}
        \includegraphics[width=\textwidth, keepaspectratio=true]{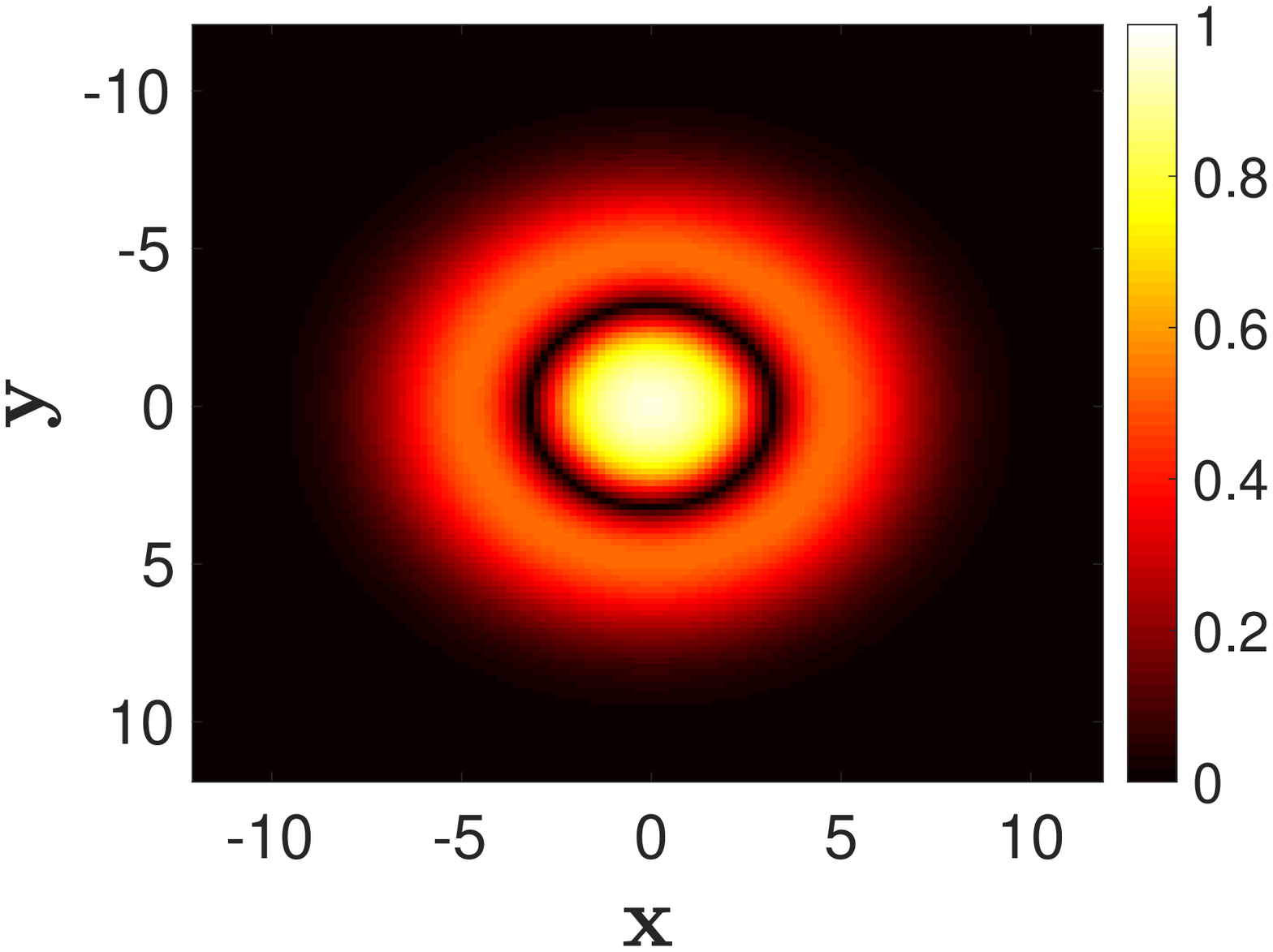}
        \caption*{(f) Br6:  $|2,0\rangle + |0,2\rangle$}
    \end{subfigure}
    \begin{subfigure}[c]{0.3\textwidth}
        \includegraphics[width=\textwidth, keepaspectratio=true]{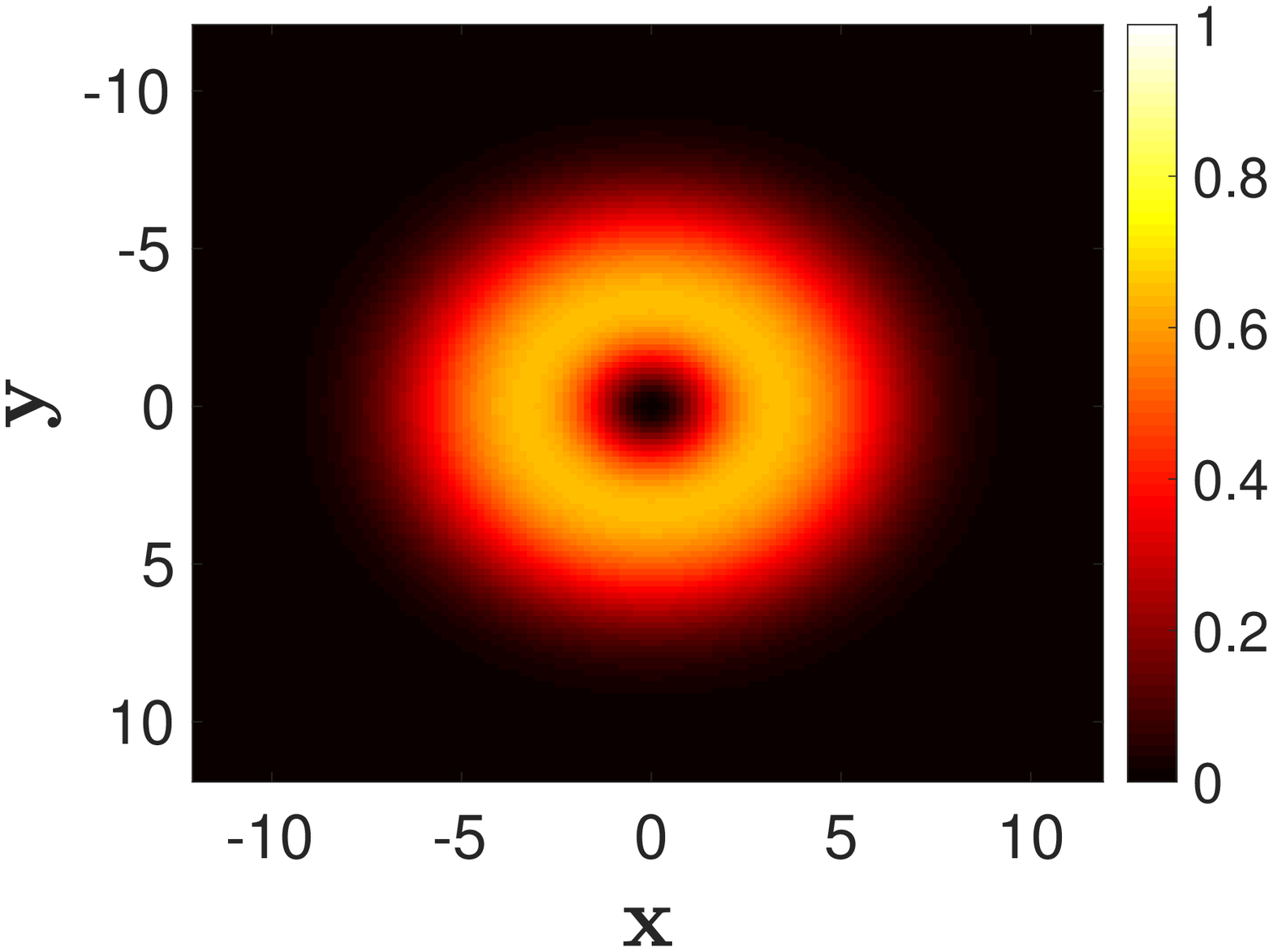}
        \caption*{(g) Br7: $|2,0\rangle - |0,2\rangle + 2i |1,1\rangle$}
    \end{subfigure}
    \begin{subfigure}[c]{0.3\textwidth}
        \includegraphics[width=\textwidth, keepaspectratio=true]{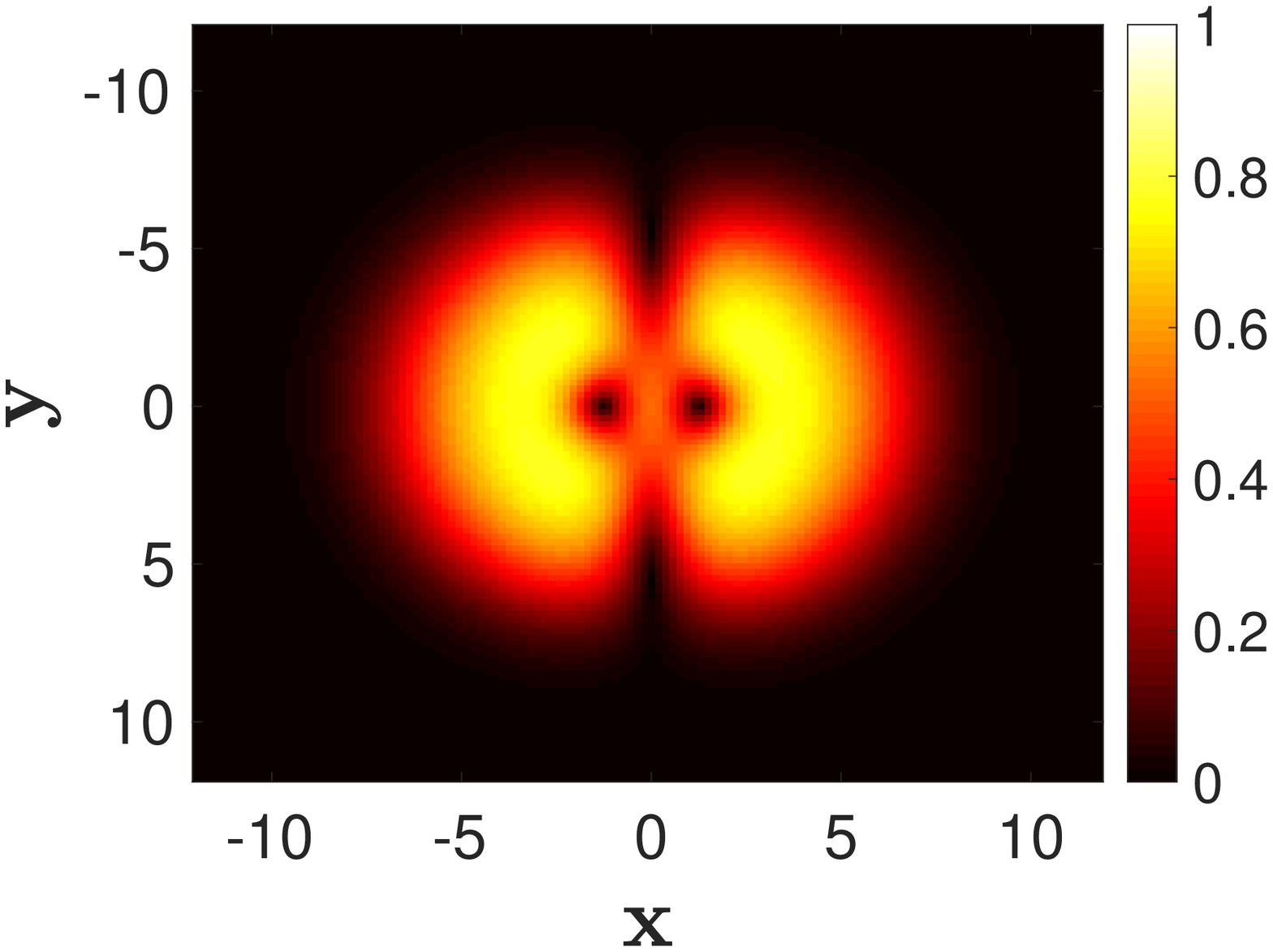}
        \caption*{(h) Br8:  $|2,0\rangle + i |1,1\rangle$}
    \end{subfigure}
    \caption{Typical examples of the different branches
      of solutions, not only the ground state one (Br1), but also
      excited ones such as the planar dark soliton (Br2), the
      single charge vortex (Br3) and so on that one can converge
    to using the SCTN method.}\label{2D_Excited}
\end{figure}

     Using these as an initial guess, we construct a partial bifurcation diagram starting at the $\mu$ values $0.2, 0.4, 0.6$. After ASCTN converged, we then increased $P$ by $\Delta P=0.5$. Fig.~\ref{fig:2D_Bif}(a) shows the corresponding bifurcation diagram and Fig.~\ref{fig:2D_Bif}(b) shows the iteration count.
     There are eight branches in total. Fig.~\ref{2D_Excited} shows plots of a selected point within each branch, as well as the relationship between the branch and the eigenstates of the associated linear limit; considerably more
     detail on the latter subject has been provided recently in~\cite{10},
     so we don't focus on the latter topic further here. 

To go from one point on a branch to the next, Fig.~\ref{fig:2D_Bif}(b) shows that it took around 100 iterations for five of the eight branches. Branches 2, 4, and 8 on the other hand consistently converged at a far higher iteration count. Branches 4 and 8 in particular took several thousand iterations to initially converge, but then settled down to around 350 for higher $P$ values. It's not clear to us why some of these converged quickly while others converged slowly.
The only thing worth mentioning is that some of these branch solutions become
unstable (with respect to time in the time-dependent NLS) already for small values of $\mu$ and progressively more so as
$\mu$ increases. 
We do want to remark however that we did not try to adaptively choose the parameters; in fact, we used the same parameter values to continue all of the branches. 

Returning to ASCTN itself, we need to mention two details. The first is that
since some of these solutions are complex both the steady
state equation and the Jacobian as written in Eq.~\eqref{ASCTN} are not accurate (the 1D equations had only real solutions). Some care needs to be
taken to find the derivative of the nonlinear term as it is not holomorphic i.e. $\frac{d}{d\psi}[ |\psi^2|\psi]$ does not exist. Instead one could split the equation itself into real and imaginary parts and then try to apply the method to a vector equation. However, we found it easier to just calculate the (real) derivative of the nonlinear term and then plug it back into \eqref{ASCTN}. Namely, letting $\psi=\psi_1 + i \psi_2$ and $H=H_1 + i H_2$, we have the directional derivative
\begin{IEEEeqnarray*}{rcl}
d(|\psi|^2 \psi)H &=& \bigg[(2\psi_1^2 + |\psi|^2)H_1 + 2\psi_1\psi_2 H_2 \bigg] + i\bigg[2\psi_1 \psi_2 H_1 + (2\psi_2^2 + |\psi|^2)H_2\bigg] \\ 
\end{IEEEeqnarray*}
where instead of writing it as a two-component vector we identified it with a complex number.

The second is that gradient restarting only applies to real functions i.e. $\langle \nabla F(\psi) , \dot{\psi} \rangle > 0$ only makes sense for real inputs. One way around this problem is to identify the given complex functions with real vector functions (under the natural identification) and then apply gradient restarting to the latter. However, recalling the identity $|u| |v| \cos \theta = Re(\langle u , v \rangle)$ in a complex inner product space, we propose the equivalent restarting scheme
\begin{equation}
Re(\langle \nabla F(\psi) , \dot{\psi} \rangle) > 0 
\end{equation}

\noindent
which works for complex functions and reduces to the former scheme when the functions are real. 

\section{Discussion and Future Work}  
In this work we have developed a collection of twists on current methods for computing both ground
and excited, and in principle both stable and unstable, stationary states of nonlinear wave equations. The following diagram summarizes the techniques used in the paper, when attempting to solve $F(u)=0$ as the stationary problem
originating from a nonlinear wave equation; the linear part of
$F$ is implicitly assumed in what follows to bear a negative Laplacian,
as it typically does for Schr{\"o}dinger type operators. 

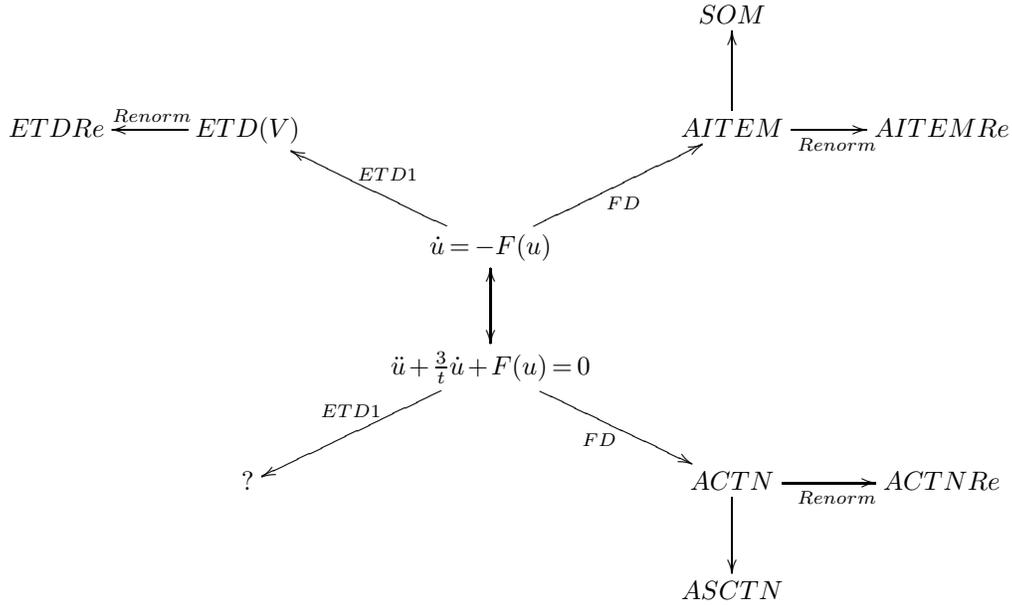
\begin{figure}
\begin{displaymath}
\xymatrixrowsep{1cm}
\xymatrixcolsep{1cm}
    \xymatrix{
      & &   &  SOM & \\
       ETDRe& \ar[l]_{Renorm}ETD(V)  &   &  AITEM\ar[r]_{Renorm}\ar[u] & AITEMRe\\
       & & \dot{u}=-F(u)\ar[ul]_{ETD1} \ar[ur]_{FD} \ar[d]& &\\
        &&  \ddot{u} + \frac{3}{t}\dot{u} + F(u) = 0\ar[dl]_{ETD1} \ar[dr]_{FD} \ar[u]&& \\
         &? & & ACTN \ar[r]_{Renorm}\ar[d]& ACTNRe\\
         &  &   &  ASCTN & \\
        }
\end{displaymath}
\caption{Schematic showing the relationships of the various methods appearing in the paper. Here FD denotes the finite difference discretization of time used to obtain the specific schemes AITEM and ACTN. Finally, the question mark represents a possible (not obtained here) scheme in which the time discretization of CTN is done via exponential time-differencing methods.}
\end{figure}

Exponential time differencing methods, given their inherent preconditioning,
are a cheap and efficient alternative to finite-difference approaches.
Traditionally, the Laplacian has been considered as the linear part in the
associated Duhamel formula; however, we have shown that there may be
advantages in considering the term bearing the potential
as the linear part instead. The future possibility of
an explicit preconditioner for ETD methods may also be of interest.

Given a constrained optimization problem and an associated
iterative procedure, we have outlined how to apply renormalization
(see the details in the Appendix) so that the constraint will be accounted
for, at least in principle. It is certainly worthwhile to explore
further how well these methods compare with other constrained optimization techniques, as well as proofs of convergence and convergence rates. In particular, in the examples above we saw that the renormalized methods were able to converge to unstable stationary states; a natural question is to what extent
can renormalized methods be engineered to converge to
(potentially even arbitrary) unstable states.

Our chief interest in this contribution, however, was to introduce and
explore the continuous time Nesterov method as applied to PDEs,
especially focusing on the elliptic, nonlinear, rich examples stemming
from the steady state problem of the nonlinear Schr{\"o}dinger
equation. For finding ground states, the examples considered imply that
accelerated continuous time Nesterov schemes generically converge linearly and
are quite competitive with other linearly converging methods; one possible
future direction of work could be devoted to establishing the linear
nature of the convergence under certain conditions. We have also shown that
a squared operator variant of such a method
will converge to excited states and the examples also
imply it has a linear convergence rate; the proof of such a feature is once
again an open problem. Developing an exponential time differencing scheme which is compatible with Nesterov type (continuous time) iterations might
provide an especially efficient way of seeking such standing waves.

On the other hand, comparing these classes of methods with Newton type
methods, or quasi-Newton ones, involving Jacobian evaluations, but also
accounting for sparsity features etc., and doing so for both one- and
multi-dimensional problems would naturally be of substantial interest.
Eventually, extending such techniques beyond steady states to periodic
orbits and limit cycles would also constitute an important step of
wide appeal to a broad and diverse array of problems.

\vspace{5mm}

{\it Acknowledgments.} PGK gratefully acknowledges support from
NSF-PHY-1602994, the Alexander von Humboldt Foundation and
the Stavros Niarchos Foundation via the Greek Diaspora
Fellowship Program. 


\newpage
\section*{Appendix}
\subsection{Renormalized Methods}
Inspired by Spectral Renormalization, we show in this section how to renormalize any iterative procedure.  Suppose 
\[x_{n+1}=F(x_n)\]
is some iteration method and 
\[P=\int |x_n|^2\]
is a constraint. One way to "enforce" this constraint when $P$ is unknown, is to introduce a renormalization constant $\lambda$ by letting
\[x = \lambda w \]
If we assume $x$ is the true fixed point, then plugging this into the iteration we get
\[\lambda w = F(\lambda w)\]
We then have that $\lambda$ solves the algebraic equation
\[0=\int w^*(\lambda w) - \int w^*F(\lambda w)\]
Now, define 
\[g(\lambda_n, w_n) = \int w_n^*(\lambda_n w_n) - \int w_n^*F(\lambda_n w_n)\]
Then we perform the new iteration
\begin{IEEEeqnarray*}{rcl}
0 &\;=\;& g(\lambda_n, w_n)  \\[.2cm]
w_{n+1} &\;=\;& \frac{1}{\lambda_n} F(\lambda_n w_n) \\[.2cm]
\end{IEEEeqnarray*}
where $\lambda_n$ is found by solving the first equation.

For concreteness, we show AITEMRe applied to the NLS equation with cubic nonlinearity:
\begin{IEEEeqnarray}{rcl}
M&\; = \;& c-\nabla^2 \nonumber \\[.2cm]
\lambda_n^2 &=& -\frac{\int \phi_n ^*(\nabla ^2 \phi_n - V(x) \phi_n + \mu \phi_n)}{\int \phi_n^*(\sigma |\phi_n |^2\phi_n)} \nonumber \\[.2cm]
\phi_{n+1} & \; = \; & \phi_n - M^{-1}(\nabla ^2 \phi_n - V(x) \phi_n + \sigma |\lambda_n \phi_n |^2\phi_n +\mu \phi_n) \Delta t
\label{AITEMRE}
\end{IEEEeqnarray}
where we have used the relationship $\psi_n = \lambda_n \phi_n$.\\

\lstset{language=Matlab,%
    basicstyle=\small, %
    breaklines=true,%
    morekeywords={matlab2tikz},
    keywordstyle=\color{blue},%
    morekeywords=[2]{1}, keywordstyle=[2]{\color{black}},
    identifierstyle=\color{black},%
    stringstyle=\color{mylilas},
    commentstyle=\color{mygreen},%
    showstringspaces=false,
    numbers=left,%
    numberstyle={\tiny \color{black}},
    numbersep=6pt, 
    emph=[1]{for,end,break},emphstyle=[1]\color{red}, 
}

\newpage
\subsection{Matlab Code}
The following is a Matlab code for ACTN applied to the 2D NLS equation
$\nabla^2 \psi - V(x) \psi + \sigma |\psi|^2 \psi + \mu \psi =0$.
\begin{verbatim}
%DEFINE VARIABLES AND PARAMETERS
L=12; m=2^7; sig=-1; dx=2*L/(m); x=(-L:dx:L-dx)'; [Xg,Yg]=ndgrid(x,x);
k=[0:m/2-1 -m/2:-1]'; k=(pi/L).*k; [xi, eta]= ndgrid(k,k);
Lap=-xi.^2-eta.^2; V= @(x,y) .02*(x.^2 + y.^2); Vx=V(Xg,Yg);
%INITIAL NORMALIZATION
XInt=@(x,y) exp(-x.^2 - y.^2);  Xi=XInt(Xg,Yg); N=17; 
Xi=Xi*(N/(sum(sum((conj(Xi).*Xi)))*dx^2))^(1/2);
%INITIALIZE ITERATE
X=Xi; X0hat=fftn(X); X1hat=X0hat; FXhat=fftn(-Vx.*X + sig*(abs(X).^2).*X);
%ITERATION PARAMETERS
dt=1; c=2; Restart=5;
ITER=1000; tol=10^-10; 
jj=0; ii=0; i=0; e=1;

while e>tol && i < ITER
    i=i+1; ii=ii+1; jj=jj+1;
    %CALCULATE MU   
    mu=-sum(sum(conj(Lap.*X1hat + FXhat).*(((1./(c-Lap)).*X1hat))))/...
        sum(sum(conj(X1hat).*(((1./(c-Lap)).*X1hat))));
    mu=real(mu);
    %ITERATION
    X2hat= (2-3/ii).*X1hat + dt^2*((1./(c-Lap))).*...
        (Lap.*X1hat + FXhat + mu*X1hat) - (1-3/ii)*X0hat;
    X2=ifftn(X2hat);
    %NORMALIZATION
    amp=(N/(sum(sum((conj(X2).*X2)))*dx^2))^(1/2);
    X2=X2*amp; X2hat=X2hat*amp;
    %RESIDUAL ERROR
     FXhat=fftn(-Vx.*X2 + sig*(abs(X2).^2).*X2);        
     e=sqrt((dx^2)/(m^2)*sum(sum((FXhat+mu*X2hat + Lap.*X2hat).*...
         conj(FXhat+mu*X2hat + Lap.*X2hat))));
    %GRADIENT RESTART
    if  sum(sum((Lap.*X1hat + FXhat + mu*X1hat).*...
            conj(X2hat -X1hat)))> 0 && ii> Restart
        ii=1;
    end
    X0hat=X1hat; X1hat=X2hat;   
end

surf(Xg,Yg,X2)





\end{verbatim}

\end{document}